\documentclass[aps,prl,twocolumn,superscriptaddress]{revtex4-1}
\usepackage[utf8]{inputenc}
\usepackage{amsmath,amssymb}
\usepackage[english]{babel}
\usepackage{color}
\usepackage{tikz}
\usepackage{graphics}
\usepackage{multirow}
\usepackage{siunitx}
\usepackage{xr}
\usepackage{lineno}
\usetikzlibrary{shapes.geometric, arrows}

\def\s{_{\mathrm{s}}}
\def\n{n}

\def\br{\mathbf{r}}

\begin{document}

\title{By-passing the Kohn-Sham equations with machine learning}

\author{Felix Brockherde}
\affiliation{Machine Learning Group, Technische Universität Berlin, Marchstr. 23, 10587 Berlin, Germany}
\affiliation{Max-Planck-Institut für Mikrostrukturphysik, Weinberg 2, 06120 Halle, Germany}
\author{Leslie Vogt}
\affiliation{Department of Chemistry, New York University, New York, NY 10003, USA}
\author{Li Li}
\affiliation{Departments of Physics and Astronomy, University of California, Irvine, CA 92697, USA}
\author{Mark E. Tuckerman}
\affiliation{Department of Chemistry, New York University, New York, NY 10003, USA}
\affiliation{Courant Institute of Mathematical Science, New York University, New York, NY 10003, USA}
\affiliation{NYU-ECNU Center for Computational Chemistry at NYU Shanghai, 3663 Zhongshan Road North, Shanghai 200062, China}
\author{Kieron Burke}
\thanks{to whom correspondence should be addressed.}
\affiliation{Departments of Chemistry, University of California, Irvine, CA 92697, USA}
\affiliation{Departments of Physics and Astronomy, University of California, Irvine, CA 92697, USA}
\author{Klaus-Robert Müller}
\thanks{to whom correspondence should be addressed.}
\affiliation{Machine Learning Group, Technische Universität Berlin, Marchstr. 23, 10587 Berlin, Germany}
\affiliation{Department of Brain and Cognitive Engineering, Korea University, Anam-dong, Seongbuk-gu, Seoul 136--713, Republic of Korea}
\affiliation{Max Planck Institute for Informatics, Stuhlsatzenhausweg, 66123 Saarbrücken, Germany}

\date{\today}

\begin{abstract}
Last year, at least 30,000 scientific papers used the Kohn-Sham
scheme of density functional theory to solve electronic structure
problems in a wide variety of scientific fields, ranging from
materials science to biochemistry to astrophysics.
Machine learning holds the promise of learning the kinetic
energy functional via examples,
by-passing the need to solve the Kohn-Sham equations.
This should yield substantial
savings in computer time, allowing either larger systems or longer time-scales
to be tackled, but attempts to machine-learn this functional have been limited by
the need to find its derivative.
The present work overcomes this difficulty by directly learning
the density-potential and energy-density maps for test systems and various molecules.
Both improved accuracy and lower computational cost with this method
are demonstrated by reproducing DFT energies for a range of molecular
geometries generated during molecular dynamics simulations.
Moreover, the methodology could be applied directly to quantum chemical calculations, allowing
construction of density functionals of quantum-chemical accuracy.
\end{abstract}

\maketitle

\section{Introduction}

Kohn-Sham density functional theory\citep{KS65} is now enormously popular as an electronic structure method in a wide variety of fields\citep{PGB15}.  Useful accuracy is achieved with standard exchange-correlation approximations, such as generalized gradient approximations\citep{P86} and hybrids\citep{B93}.  Such calculations are playing a key role in the materials genome initiative\citep{MGI}, at least for weakly correlated materials\citep{jain2011}.

There has also been a recent spike of interest in applying machine learning (ML) methods in the physical sciences\citep{PHSR12,MP13,FDP14,rupp2012fast,hautier2010finding}. The majority of these applications involve predicting properties of molecules or materials from large databases of KS-DFT calculations\citep{hansen2013,schutt2014,HBRP15,Schutt2017}.
A few applications involve finding potential energy surfaces within MD simulations\citep{BehlerParinello2007,SekoTakahashiTanaka2014,LKD2015,Chmiela2017}.
Fewer still have focussed on finding the functionals of DFT as a method of performing KS electronic structure calculations without solving the KS equations\citep{snyder2012,snyder2013orbitalfree,li2015,li2016}.
If such attempts could be made practical, the possible speed-up in repeated DFT calculations of similar species, such as occur in ab initio MD simulations, is enormous.

\begin{figure*}
  \includegraphics[width=2\columnwidth]{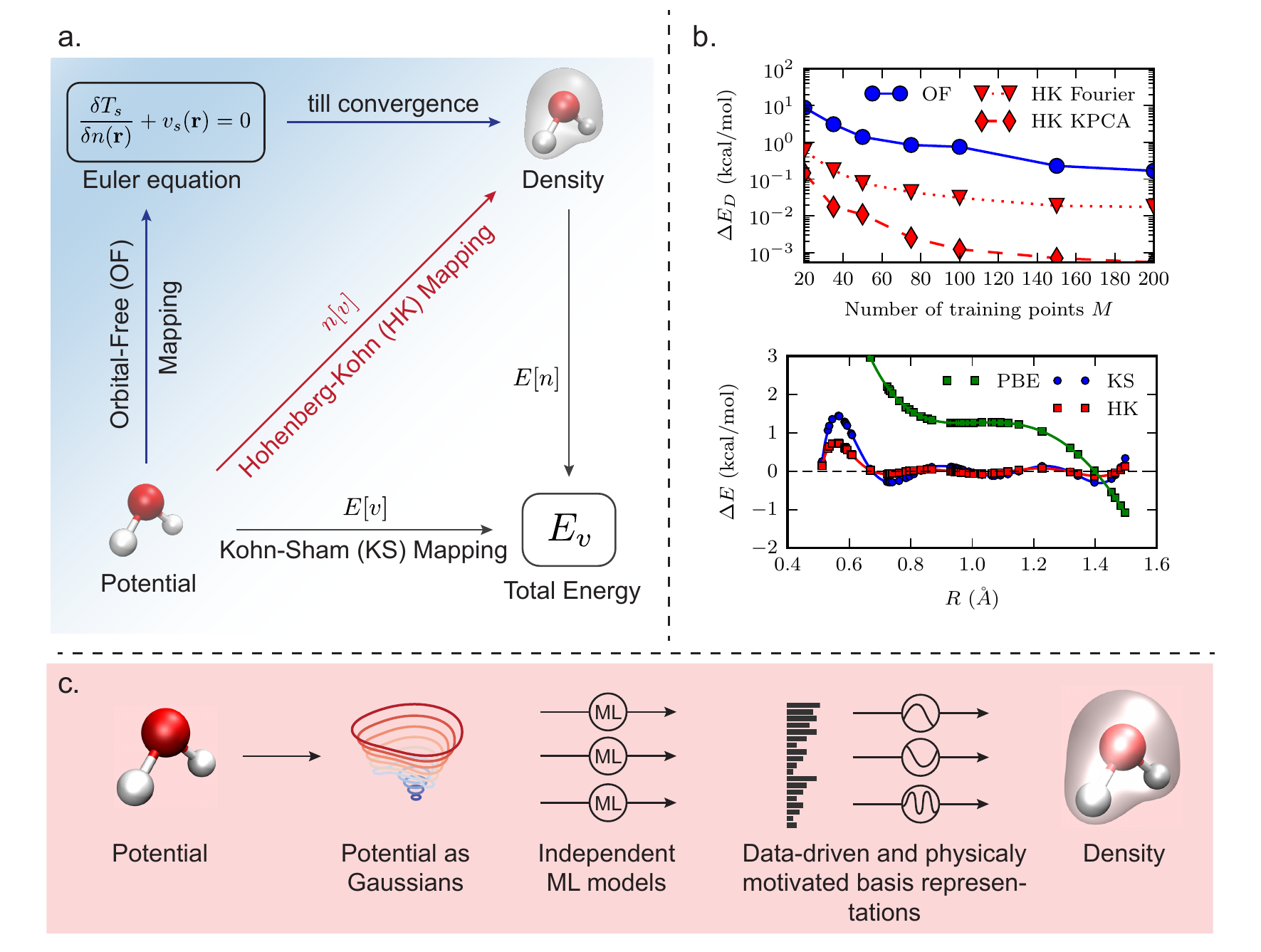}
  \caption{\label{fig:overview}
    \textbf{a.} Mappings used in this paper. The bottom arrow represents $E[v]$, a conventional electronic structure calculation, i.e., KS-DFT\@. The ground state energy is found by solving KS equations given the external potential, $v$. $E[\n]$ is the total energy density functional. The red arrow is the HK map $n[v]$ from external potential to its ground state density.
    \textbf{b} top. How the energy error depends on $M$ for ML-OF and ML-HK with different basis sets for the 1-D problem.
    \textbf{b} bottom. Errors of the PBE energies (relative to exact values) and the ML maps (relative to PBE) as a function of interatomic spacing, $R$, for $\mathsf{H_2}$ with $M~=~7$.
    \textbf{c}. How our Machine Learning Hohenberg-Kohn (ML-HK) map makes predictions.
The molecular geometry is represented by Gaussians; many independent Kernel Ridge Regression models predict each basis coefficient of the density. We analyze the performance of data-driven (ML) and common physical basis representations for the electron density.
  }
\end{figure*}

A key difficulty has been the need to extract the functional derivative of the non-interacting kinetic energy.  The non-interacting kinetic energy functional $T\s[\n]$ of the density $\n$ is used in {\em two\/} distinct ways in a KS calculation\citep{KS65}, as illustrated in Fig.~\ref{fig:overview}:
(i) its functional derivative is used in the Euler equation which is solved in the self-consistent cycle and (ii) when self-consistency is reached, the ground-state energy of the system is calculated by $E[\n]$, an Orbital-Free (OF) mapping. The solution of the KS equations performs both tasks exactly. Early results on simple model systems showed that machine learning could provide highly accurate values for $T\s[\n]$ with only modest amounts of training\citep{snyder2012}, but that the corresponding functional derivatives are too noisy to yield sufficiently accurate results to (i).  Subsequent schemes overcome this difficulty in various ways, but typically lose a factor of 10 or more in accuracy\citep{li2015}, and their computational cost can increase dramatically with system complexity.

Here we present an alternative ML approach, in which we replaced the Euler equation by directly learning the Hohenberg-Kohn (HK) map $v(\br)\to\n(\br)$ (red line in Fig.~\ref{fig:overview}a) from the one-body potential of the system of interest to the interacting ground-state density, i.e.~we establish an ML-HK map. We show that this map can be learned at a much more modest cost than either previous ML approaches to find the functional and its derivative (ML-OF) or direct attempts to model the energy as a functional of $v(\br)$ (ML-KS). Furthermore we show that it can immediately be applied to molecular calculations, by calculating the energies of small molecules over a range of conformers. Moreover, since we have already implemented this approach with a standard quantum chemical code (Quantum Espresso\citep{QE2009}) using a standard DFT approximation (PBE), this can now be tried on much larger scales.

The ML-HK map reflects the underlying computational approach used to generate a particular electron density, but is not restricted to any given electronic structure method.  Many molecular properties, not only the energy, are dependent on the electron density, making the ML-HK map more versatile than a direct ML-KS mapping.
We also establish that densities can be learned with sufficient accuracy to distinguish between different DFT functionals, providing a route to future functional development by generating precise densities for a range of molecules and conformations.

\section{Results}

We will first outline theoretical results, most prominently the ML-HK map, and then illustrate the approach with simulations of 1-D systems and 3-D molecules.

\subsection{ML-Hohenberg-Kohn map}
Previous results show that for an ML-OF approach, the accuracy of ML KS kinetic energy models $T\s^\text{ML}[\n]$ improve rapidly with the amount of data. But minimizing the total energy via gradient descent requires the calculation of the \textit{gradient} of the kinetic energy model $T\s^\text{ML}$ (see Fig.~\ref{fig:overview}). Calculating this gradient is challenging. Due to the data driven nature of, e.g., kernel models, the machine-learned kinetic energy functional has no information in directions that point outside the data manifold\citep{snyder2013kernels}. This heavily influences the gradient to an extent that it becomes unusable without further processing\citep{snyder2012}. There have been several suggestions to remedy this problem but all of them share a significant loss in accuracy compared to $T\s[\n]$\citep{snyder2013orbitalfree,snyder2015nonlinear,li2015}.

However, we propose an interesting alternative to gradients and the ML-OF approach. Recently, it has been shown that the Hohenberg-Kohn map for the density as a functional of the potential can be approximated extremely accurately using semiclassical expressions\citep{RLCE14}. Such expressions do not require the solution of any differential equation, and become more accurate as the number of particles increases. Errors can be negligible even for just 2 distinct occupied orbitals.

Inspired by this success, we suggest to circumvent the kinetic energy gradient and directly train a multivariate machine learning model. We name this the ML-Hohenberg-Kohn (ML-HK) map:

\begin{align}
  n^\text{ML}[v](x) = \sum_{i=1}^M \beta_{i}(x) k(v, v_i).
\end{align}
Here, each density grid point is associated with a group of model weights $\boldsymbol{\beta}$.
Training requires solving an optimization problem for each density grid point. While this is possible in 1-D, it rapidly becomes intractable in 3-D, since the number of grid points grows cubically.

The use of a basis representation for the densities, as in

\begin{align}
  n^\text{ML}[v](x) = \sum_{l=1}^L u^{(l)}[v]\phi_l(x),
\end{align}
renders the problem tractable even for 3-D.
A machine learning model that predicts the basis function coefficients $u^{(l)}[v]$ instead of the grid points is then formulated.

Predicting the basis function coefficients not only makes the machine learning model efficient and allows the extension of the approach to 3-D but also permits regularization, e.g.~to smooth the predicted densities by removing the high frequency basis functions for example, or to further regularize the machine learning model complexity for specific basis functions.

For orthogonal basis functions, the machine learning model reduces to several independent regression models and admits an analytical solution analogous to Kernel Ridge Regression (see supplement Eq.~\ref{sup-eq:krr-analytical}):

\begin{align} \label{eq:ml-hk-analytical}
\boldsymbol{\beta}^{(l)} = {\left(\mathbf{K}_{\sigma^{(l)}} + \lambda^{(l)} \mathbf{I}\right)}^{-1} \mathbf{u}^{(l)}, \quad l = 1, \dots, L.
\end{align}
Here, for each basis function coefficient, $\lambda^{(l)}$ are regularization parameters and $K_{\sigma^{(l)}}$ is a Gaussian kernel with kernel width $\sigma^{(l)}$.
The $\lambda^{(l)}$ and $\sigma^{(l)}$ can be chosen individually for each basis function via independent cross-validation (see \citep{muller2001introduction,hansen2013}).
This ML-HK model avoids prior gradient descent procedures and with it the necessity to ``de-noise'' the gradients.
Due to the independence of Eq.~\ref{eq:ml-hk-analytical} for each $l$, the solution scales nicely.

\subsection{Functional and Density driven error}

How can the performance of the ML-HK map be measured? It has recently been shown how to separate out
the effect of the error in the functional $F$ and the error in the
density $\n(\br)$ on the resulting error in the total energy of any
approximate, self-consistent DFT calculation\citep{KSB13}. Let $\tilde{F}$ be an
approximation of the many body functional $F$, and $\tilde{\n}(\br)$ the approximate ground-state
density when $\tilde F$ is used in the Euler equation. Defining
$\tilde E[\n]=\tilde F[\n] + \int d^3r \n(\br)v(\br)$ yields
\begin{equation}
  \Delta E = \tilde E[\tilde\n]- E[\n] = \Delta E_F + \Delta E_D
\end{equation}
where $\Delta E_F = \tilde{F}[n] - F[n]$ is the functional-driven error,
while $\Delta E_D = \tilde E[\tilde\n] - \tilde{E}[\n]$ is the density-driven error.
In most DFT calculations, $\Delta E$ is dominated by $\Delta E_F$.
The standard DFT approximations can, in some specific cases,
produce abnormally large density errors that dominate the
total error.  In such situations, using a more accurate density
can greatly improve the result
\citep{KSB13, KSB14, KPSS15}.
We will use these definitions to measure the accuracy of the ML-HK map.

\subsection{1-D potentials}

The following results demonstrate how much more accurate ML is when applied directly to the HK map.
The box problem originally introduced in \citet{snyder2012} is used to illustrate the principle.
Random potentials consisting of three Gaussian dips were generated inside
a hard-wall box of length 1 (atomic units), and the Schrödinger equation
for one electron was solved extremely precisely.
Up to 200 cases were used to train an ML model $T\s^\mathrm{ML}[n]$ for the non-interacting kinetic energy functional $T\s[\n]$ via Kernel Ridge Regression (for details, see supplement).

\begin{table*}[htb]
\begin{ruledtabular}
  \begin{tabular}{r|cccccc|cccccc|cccc}
  & \multicolumn{6}{c|}{ML-OF} & \multicolumn{6}{c|}{ML-HK (grid)} & \multicolumn{4}{c}{ML-HK (other)}\\
  \noalign{\smallskip}
  \cline{2-7}\cline{8-13}\cline{14-17}
  \noalign{\smallskip}
  & \multicolumn{2}{c}{$\Delta E$} & \multicolumn{2}{c}{$\Delta E_F$} & \multicolumn{2}{c|}{$\Delta E_D$} &
      \multicolumn{2}{c}{$\Delta E$} & \multicolumn{2}{c}{$\Delta E_D$} & \multicolumn{2}{c|}{$\Delta E_D^{\mathrm{ML}}$} & \multicolumn{2}{c}{$\Delta E_D$ (Fourier)} & \multicolumn{2}{c|}{$\Delta E_D$ (KPCA)} \\
  \noalign{\smallskip}
  \cline{2-3}\cline{4-5}\cline{6-7}\cline{8-9}\cline{10-11}\cline{12-13}\cline{14-15}\cline{16-17}
  \noalign{\smallskip}
  $M$ & MAE  &  max  &  MAE  &  max  &  MAE  &  max  &  MAE  &  max  &  MAE  &  max  &  MAE  &  max  &  MAE  &  max  &  MAE  &  max\\
  \noalign{\smallskip}
  \colrule
  \noalign{\smallskip}
   20  &   7.7 &    47 &   7.7 &    60 &   8.8 &    87 &   3.5 &    27 &  0.76 &   8.9 &   9.7 &    70 &  0.58 &     8 &  0.15 &   2.9\\
   50  &   1.6 &    30 &   1.3 &   7.3 &   1.4 &    31 &   1.2 &   7.1 & 0.079 &  0.92 &  0.27 &   2.4 & 0.078 &  0.91 & 0.011 &  0.17\\
  100  &  0.74 &    17 &   0.2 &   2.6 &  0.75 &    17 &  0.19 &   2.1 & 0.027 &  0.43 &  0.18 &   2.4 & 0.031 &  0.42 & 0.0012 & 0.028\\
  200  &  0.17 &   2.9 & 0.039 &   0.6 &  0.17 &   2.9 & 0.042 &  0.59 & 0.0065 &  0.15 &  0.02 &  0.46 & 0.017 &  0.14 & 0.00055 & 0.015\\
  \end{tabular}
\end{ruledtabular}
\caption{\label{tab:results_1d}
Energy errors in kcal/mol for the 1-D data set for various $M$, the number of training points. For definitions, see text.
}
\end{table*}

To measure the accuracy of an approximate HK map, the analysis of the previous section is applied to the KS DFT problem.
Here $F$ is just $T\s$, the non-interacting
kinetic energy, and

\begin{align}
\Delta E_F = \tilde T\s[\n]-T\s[\n],
\end{align}
i.e., the error made in an approximate functional on the exact density.
Table~\ref{tab:results_1d} on the left gives the errors made by ML-OF for the total energy, and its different components,
when the density is found from the functional derivative. This method works by following a gradient descent of the total energy functional based on the gradient of the ML model $T\s^\mathrm{ML}$,

\begin{align}
  n^{(j+1)} = n^{(j)} - \epsilon P\left(n^{(j)}\right)\frac{\delta}{\delta n}E^\mathrm{ML}(n^{(j)}),
\end{align}
where $\epsilon$ is a small number and $P(n^{(j)})$ is a localized PCA projection to de-noise the gradient.
Here and for all further 1-D results we use
\begin{align}
  E^\mathrm{ML}[n] = T\s^\mathrm{ML}[n] + \int \mathrm{d} x\,  n(x)\, v(x).
\end{align}
The density-driven contribution to the error $\Delta E_D$, which we calculate exactly here using the von Weizsäcker kinetic energy\cite{DG90} is always comparable to, or greater
than, the functional-driven error $\Delta E_F$, due to the poor quality of the ML functional derivative\citep{snyder2012}.
The calculation is abnormal, and
can be greatly improved by using a more accurate density from a finer grid.  As the number of training points $M$ grows,
the error becomes completely dominated by the error in the density.
This shows that the largest source of error is in using the ML
approximation of $T\s$ to find the density by solving the Euler equation.

The next set of columns analyzes the ML-HK approach, using a grid basis.  The left-most of these columns shows the energy error we obtain by utilizing the ML-HK map:

\begin{align}
  \Delta E = |E^\mathrm{ML}[n^\mathrm{ML}[v]] - E|.
\end{align}
Note that both ML models, $T\s^\mathrm{ML}$ and $n^\mathrm{ML}$, have been trained using the same set of $M$ training points.

The ML-HK approach is always more accurate than ML-OF, and its relative
performance improves as $M$ increases.  The next column reports the density-driven error $\Delta E_D$ which is an order-of-magnitude smaller than for ML-OF\@. Lastly, we list an estimate to the density-driven error

\begin{align} \label{eq:Delta_E_D}
  \Delta E_D^\mathrm{ML} = |E^\mathrm{ML}[n^\mathrm{ML}[v]] - E^\mathrm{ML}[n]|,
\end{align}
which uses the ML model $T\s^\mathrm{ML}$ for the kinetic energy functional in 1-D.
This proxy is generally a considerable overestimate (a factor
of 3 too large), so that the true $\Delta E_D$ is always significantly smaller.
We use it in subsequent calculations (where we cannot
calculate $T\s^\mathrm{ML}$) to
(over-)estimate the energy error due to the HK-ML map.

The last set of columns are density-driven errors for other basis sets.  Three variants of the ML-HK map were tested.
First, direct prediction of the grid coefficients: In this case,   $\mathbf{u}_i^{(l)} = n_i(x_l)$, $l=1,\dots,G$. 500 grid points were used, as in \citet{snyder2012}. This variant is tested in 1-D only; in 3-D the high dimensionality will be prohibitive.
Second, a common Fourier basis is tested. The density can be transformed efficiently via the discrete Fourier transform, using 200 Fourier basis functions in total. In 3-D these basis functions correspond to plane waves. The back-projection $\mathbf{u} \mapsto n$ to input space is simple, but although the basis functions are physically motivated, they are very general and not specifically tailored to density functions.
The performance is almost identical to the grid on average,
although maximum errors are much less.
For $M=20$, the error that originates from the basis representation starts to dominate.
This is a motivation for exploring, third, a Kernel PCA (KPCA) basis\citep{inputspacevs}. KPCA\citep{scholkopf1998nonlinear} is a popular generalization of PCA that yields basis functions that maximize variance in a higher dimensional feature space. The KPCA basis functions are data-driven and computing them requires an eigen-decomposition of the Kernel matrix. Good results are achieved with only 25 KPCA basis functions. The KPCA approach gives better results because it can take the non-linear structure in the density space into account. However, it introduces the pre-image problem: It is not trivial to project the densities from KPCA space back to their original (grid) space (see supplement). It is thus not immediately applicable to 3-D applications.

\subsection{Molecules}
We next apply the ML-HK approach to predict electron densities and energies for a series of small molecules. We test the ML models on KS-DFT results obtained using the PBE exchange-correlation functional\cite{perdew1996generalized} and atomic pseudopotentials with the projector augmented wave (PAW) method\cite{kresse1999ultrasoft,blochl1994projector} in the Quantum ESPRESSO code.\cite{giannozzi2009quantum}
Since the ML-OF approach applied in the previous section becomes prohibitively expensive in 3-D due to the poor convergence of the gradient descent procedure, we compare the ML-HK map to the ML-KS approach. This approach models the energy directly as a functional of $v(\br)$, i.e.\ it trains a model

\begin{align}
  E^{\text{ML}}[v] = \sum_{i=1}^M \alpha_i k(v_i, v)
\end{align}
via KRR (for details, see supplement).

We also apply the ML-HK map with Fourier basis functions. Instead of a $T\s^{\mathrm{ML}}[n]$ model, we learn an $E^\mathrm{ML}[n]$ model

\begin{align}
 E^\mathrm{ML}[n] = \sum_{i=1}^M \alpha_i k(n_i, n)
\end{align}
which avoids implementing the potential energy and exchange-correlation
functionals.
%

Both approaches require the characterization of the Hamiltonian by its external potential. The external (Coulomb) potential diverges for the 3-D molecules and is therefore not a good feature to measure the distance in ML\@. Instead, we use an artificial \textit{Gaussians potential} in the form of

\begin{align}
    v(\br) = \sum_{\alpha=1}^{N^a} Z_\alpha \exp \left( \frac{-\lVert \br - R_\alpha \rVert^2}{2\gamma^2} \right)
\end{align}
where $R_\alpha$ are the positions and $Z_\alpha$ are the nuclear charges of the $N^a$ atoms.
The Gaussians potential is used for the ML representation only. The width $\gamma$ is a hyper-parameter of the algorithm. The choice is arbitrary but can be cross-validated. We find good results with $\gamma = 0.2~\si{\angstrom}$.
The idea of using Gaussians to represent the external potential has been used previously\cite{bartok2010gaussian}. The Gaussians potential is discretized on a coarse grid with grid spacing $\Delta = 0.08~\si{\angstrom}$. To use the discretized potential in the Gaussian kernel, we flatten it into a vector form and thus use a tensor Frobenius norm.

Our first molecular prototype is $\mathsf{H_2}$, with the only degree of freedom, $R$, denoting the distance between the $\mathsf{H}$ atoms. A dataset of 150 geometries is created by varying $R$ between $0.5$ and $1.5~\si{\angstrom}$ (sampled uniformly).
A randomly chosen subset of 50 geometries are designated as the \textit{test set} and are unseen by the ML algorithms. These geometries are used to measure the \textit{out-of-sample} error reported below.

The remaining 100 geometries make up the \textit{grand training set}.
To evaluate the performance of the ML-KS map and the ML-HK map, subsets of varying sizes $M$ are chosen out of the grand training set to train the $E^{\text{ML}}[v]$ and $n^{\text{ML}}[v]$ models, respectively.
Because the required training subsets are so small, careful selection of a subset that covers the complete range of $R$ is necessary.
This is accomplished by selecting the $M$ training points out of the grand training set so that the $R$ values are nearly equally spaced (see supplement for details).

For practical applications, it is not necessary to run DFT calculations for the complete grand training set, only for the $M$ selected training points. Strategies for sampling the conformer space and selecting the grand training set for molecules with more degrees of freedom are explained for $\mathsf{H_2O}$ and MD simulations later on.

\begin{table*}[htb]
  \begin{minipage}[t]{0.67\textwidth}
\begin{ruledtabular}
\begin{tabular}{cr|cccc|cccccc}
\noalign{\smallskip}
& & \multicolumn{4}{c|}{ML-KS} & \multicolumn{6}{c}{ML-HK}\\
\noalign{\smallskip}
\cline{3-6}\cline{7-12}
\noalign{\smallskip}
& & \multicolumn{2}{c} {$\Delta E$} &$\Delta R_o$ & $\Delta \theta_0$
& \multicolumn{2}{c}{$\Delta E$} & \multicolumn{2}{c} {$\Delta E_D^\mathrm{ML}$} & $ \Delta R_o$ & $\Delta \theta_0$ \\
\noalign{\smallskip}
\cline{3-4}\cline{5-5}\cline{6-6}\cline{7-8}\cline{9-10}\cline{11-11}\cline{12-12}
\noalign{\smallskip}
Molecule & $M$ & MAE & max & & &MAE & max &MAE & max & & \\
\noalign{\smallskip}
\colrule
\noalign{\smallskip}
& 5 & 1.3 & 4.3 & 2.2 & --- & 0.70 & 2.9 & 0.18 & 0.54 & 1.1 & --- \\
$\mathrm{H_2}$ & 7 & 0.37 & 1.4 & 0.23 & --- & 0.17 & 0.73 & 0.054 & 0.16 & 0.19 & --- \\
& 10 & 0.080 & 0.41 & 0.23 & --- & 0.019 & 0.11 & 0.017 & 0.086 & 0.073 & --- \\
\noalign{\smallskip}
\colrule
\noalign{\smallskip}
\multirow{6}{*}{$\mathrm{H_2O}$} & 5 & 1.4 & 5.0 & 2.1 & 2.2 & 1.1 & 4.9 & 0.056 & 0.17 & 2.3 & 3.8 \\
& 10 & 0.27 & 0.93 & 0.63 & 1.9 & 0.12 & 0.39 & 0.099 & 0.59 & 0.12 & 0.38 \\
& 15 & 0.12 & 0.47 & 0.19 & 0.41 & 0.043 & 0.25 & 0.029 & 0.14 & 0.064 & 0.23 \\
& 20 & 0.015 & 0.064 & 0.043 & 0.16 & 0.0091 & 0.060 & 0.011 & 0.058 & 0.024 & 0.066 \\
\noalign{\smallskip}
\end{tabular}
\end{ruledtabular}
\end{minipage}\hfill
\begin{minipage}[c]{0.30\textwidth}
\vspace{0pt}
\caption{\label{tab:results_3d_H2}
Prediction errors on $\mathsf{H_2}$ and $\mathsf{H_2O}$ with increasing number of training points $M$ for the ML-KS and ML-HK approaches. In addition, the estimated density-driven contribution to the error for the ML-HK approach (Eq.~\ref{eq:Delta_E_D}) is given. Energies in kcal/mol, bond-lengths in pm, and angles in degrees.}
\end{minipage}
\end{table*}

The performance of the ML-KS map and ML-HK map is compared by evaluating $E^{\text{ML}}[v]$ that maps from the Gaussians potential to total energy and the combination of $n^{\text{ML}}[v]$ that maps from Gaussians potential to the ground-state density in a three-dimensional Fourier basis representation ($l=25$) and $E^{\text{ML}}[n]$ that maps from density to total energy. The prediction errors are listed in Table~\ref{tab:results_3d_H2}.

The MAE of the energy evaluated using the ML-HK map is significantly smaller than that of the ML-KS map.
This indicates that even for a 3-D system, learning the potential-density relationship via the HK map is much easier than directly learning the potential-energy relationship via the KS map.

Fig.~\ref{fig:overview}b shows the errors made by the ML-KS and the ML-HK maps.
The error of the ML-HK map is smoother than the ML-KS error and is much smaller, even for the most problematic region when $R$ is smaller than the equilibrium bond distance of $R_0=0.74$ \AA\@.
The MAE that is introduced by the PBE approximation on the $\mathsf{H_2}$ dataset is 2.3 kcal/mol (compared to exact CI calculations), i.e., well above the errors of the ML model and verifies that the error introduced by the ML-HK map is negligible for a DFT calculation.

The next molecular example is $\mathsf{H_2O}$,
parametrized with three degrees of freedom:
two bond lengths and a bond angle.
To create a conformer dataset, the optimized structure ($R_0=0.97$ \AA, $\theta_0=104.2\si{\degree}$ using PBE) is taken as a starting point. A total of 350 geometries are then generated by changing each bond length by a uniformly sampled value between $\pm 0.075$ \AA\ and varying the angle $\theta$  between $\pm 8.59$ degrees ($\pm 0.15$ rad) away from $\theta_0$
(see supplement Fig.~\ref{sup-fig:h2o-dataset} for a visualization of the sampled range).
For this molecule, the out-of-sample test set again comprises a random subset of 50 geometries, with the remaining 300
geometries used as the grand training set.
Because there are now three parameters, it is more difficult to select equidistant samples for the training subset of $M$ data points. We therefore use a K-means approach to find $M$ clusters and select the grand training set geometry closest to each cluster's center for the training subset
(see supplement for details).

\begin{figure}[htb]
\includegraphics[width=\columnwidth]{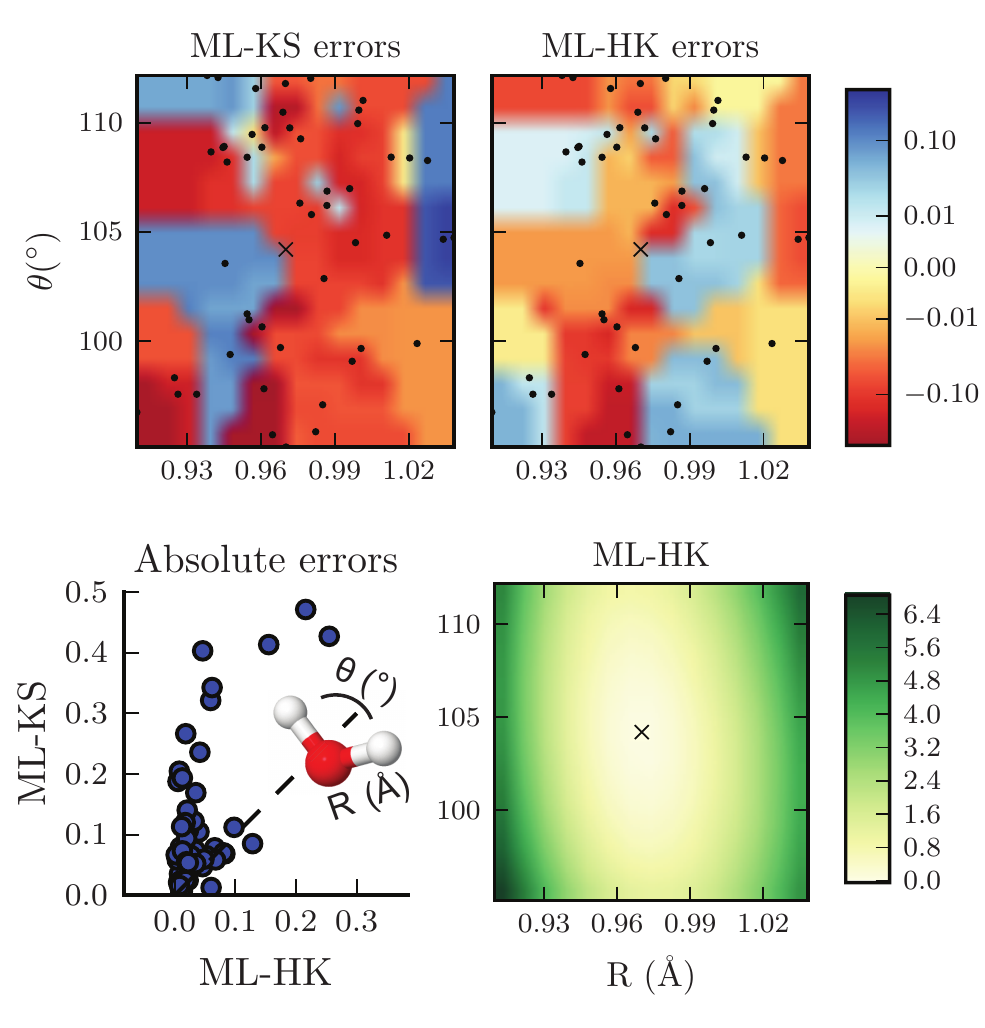}
\caption{\label{fig:H2O} \textbf{Top.} Distribution of energy errors against PBE on the $\mathsf{H_2O}$ dataset for ML-KS and ML-HK\@. The errors are plotted on a symmetric log scale with linear threshold of 0.01, using nearest neighbor interpolation from a grid scan for coloring. Black dots mark the test set geometries with averaged bond lengths. \textbf{Bottom left.} Comparison of the PBE errors made by ML-HK and ML-KS on the test set geometries. \textbf{Bottom right.} Energy landscape of the ML-HK map for symmetric geometries ($R$ versus $\theta$).  All models trained on $M=15$ training points. Energies and errors in kcal/mol. A black cross marks the PBE equilibrium position.}
\end{figure}

Models are trained as for $\mathsf{H_2}$. The results are given in Table~\ref{tab:results_3d_H2}.
As expected, the increase in degrees of freedom for $\mathsf{H_2O}$ compared to $\mathsf{H_2}$ requires a larger training set size $M$. However, even for the more complicated molecule, the ML-HK map is consistently more precise than the ML-KS map, and provides an improved potential energy surface, as shown in Fig.~\ref{fig:H2O}. With an MAE of 1.2 kcal/mol for PBE energies relative to CCSD(T) calculations for this data set, we again show that ML does not introduce a new significant source of error.

The ML maps can also be used to find the minimum energy configuration. The total energy is minimized as the geometry varies with respect to both bond lengths and angles.
For optimization, we use Powell's method\cite{P64}, which requires a starting point and an evaluation function to be minimized.
For the $\mathsf{H_2O}$ case, the search is restricted to symmetric configurations, with a random symmetric geometry used as the starting point.
Results are reported in Table~\ref{tab:results_3d_H2}.
The optimizations consistently converge to the correct minima regardless of starting point, consistent with the maps being convex, i.e., the potential energy curves are sufficiently smooth as to avoid introducing artificial local minima.

\begin{figure*}[htb]
  \includegraphics[width=\textwidth]{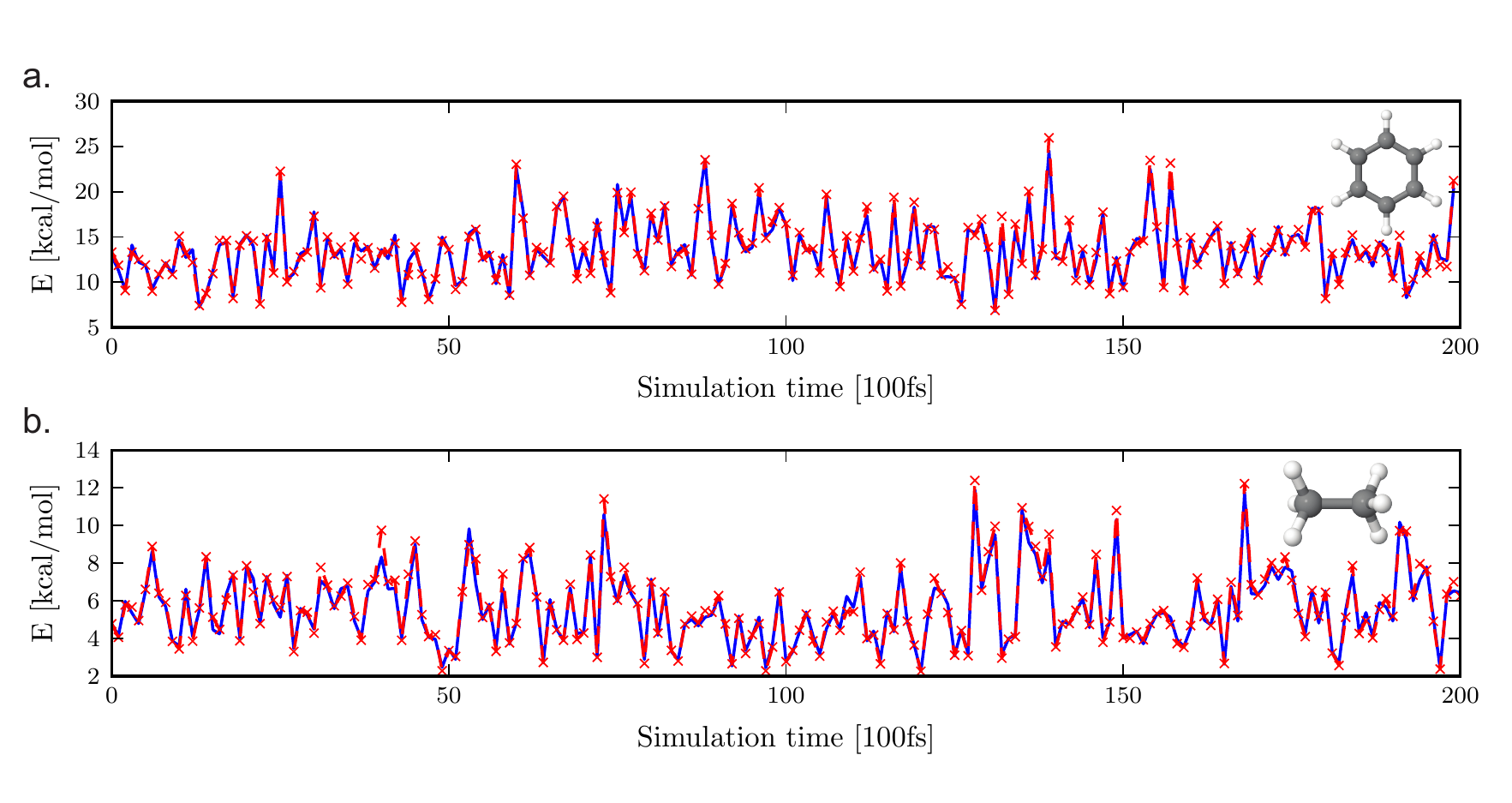}
  \caption{\label{fig:trajectories} Energy errors of ML-HK along classical MD trajectories. PBE values in blue, ML-HK values in red. \textbf{a.} A 20~ps classical trajectory of benzene. \textbf{b.} A 20~ps classical trajectory of ethane.}
\end{figure*}

For larger molecules, generating random conformers that sample the full configurational space becomes difficult.  Therefore, we next demonstrate that molecular dynamics (MD) using a classical force field can also be used to create the grand training set. As an example, we use benzene ($\mathsf{C_6H_6}$) with only small fluctuations in atomic positions out of the molecular plane. Appropriate conformers are generated via isothermal MD simulations at 300~K, 350~K, and 400~K using the General Amber Force Field (GAFF)\cite{GAFF} in the PINY\_MD package\cite{PINYMD}. Saving snapshots from the MD trajectories generates a large set of geometries that are sampled using the K-means approach to obtain 2,000 representative points for the grand training set.
Training $n^{\text{ML}}[v]$ and $E^{\text{ML}}[n]$ is performed as above by running DFT calculations on $M = 2000$ points. We find that the ML error is reduced by creating the training set from trajectories at both the target temperature and a higher temperature to increase the representation of more distorted geometries.  The final ML model is tested on 200 conformational snapshots taken from an independent MD trajectory at 300~K (see Fig.~\ref{fig:trajectories}a). The MAE of the ML-HK map for this data set using training geometries from 300~K and 350~K trajectories is only 0.37~kcal/mol for an energy range that spans more than 10~kcal/mol (see Table~\ref{tab:results_mol}).

\begin{figure*}[htb]
  \includegraphics[width=\textwidth]{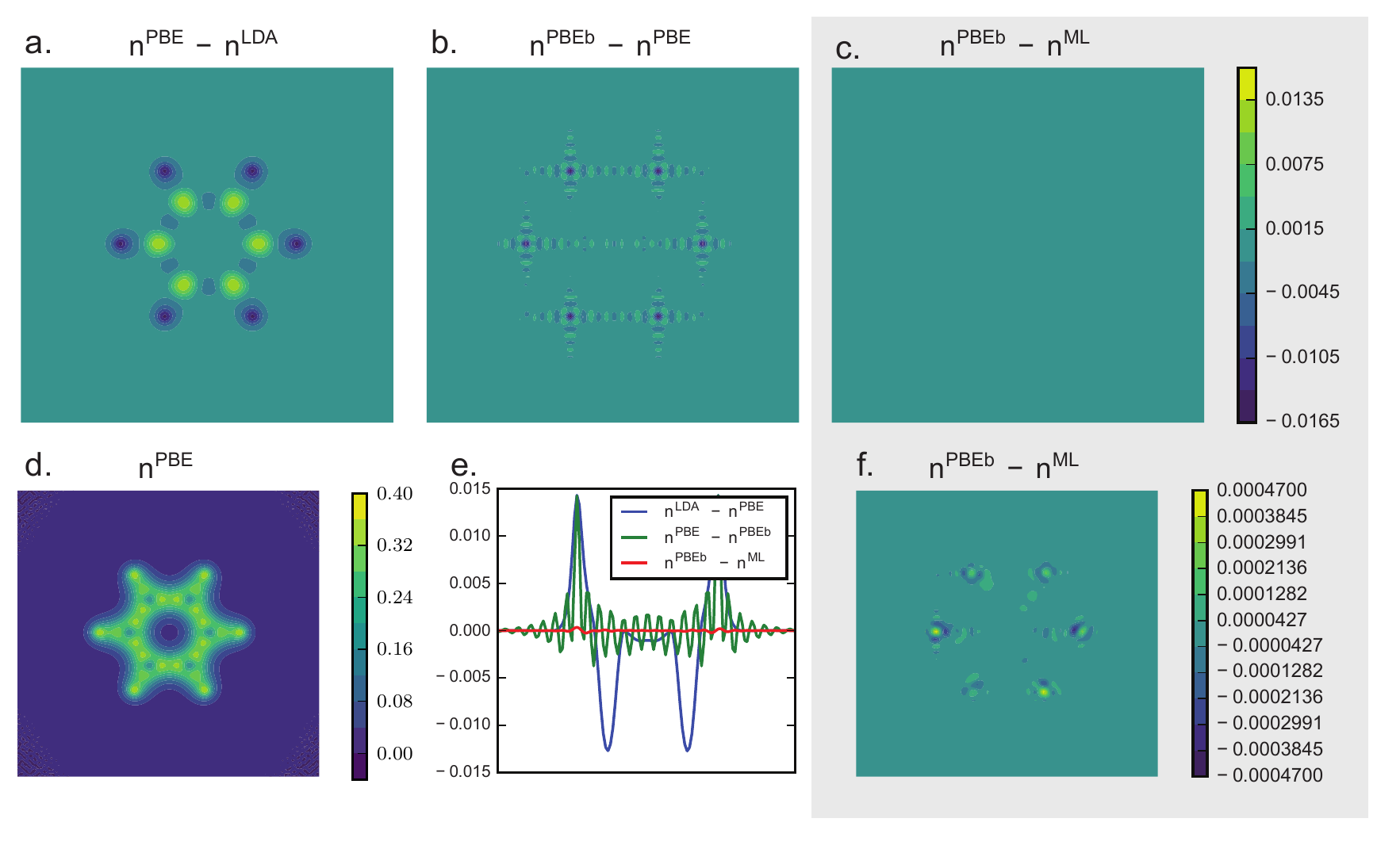}
  \caption{\label{fig:density_comparison} The precision of our density predictions using the Fourier basis for ML-HK for the molecular plane of benzene. The plots show
  $\mathbf{a.}$ the difference between the valence density of benzene when using PBE and LDA functionals at the PBE optimized geometry.
  $\mathbf{b.}$ error introduced by using the Fourier basis representation.
  $\mathbf{c.}$ error introduced by the $n^{\text{ML}}[v]$ density fitting (a.--c.\ on same color scale).
  $\mathbf{d.}$ the total PBE valence density
  $\mathbf{e.}$ the density differences along a 1-D cut of a.--c.
  $\mathbf{f.}$ the density error introduced with the ML-HK map (same data, but different scale, as in c.). }
\end{figure*}

For benzene, we further quantify the precision of the ML-HK map in reproducing PBE densities. In Fig.~\ref{fig:density_comparison}, it is clear that the errors in the Fourier basis representation are larger than the errors introduced by the ML-HK map by two orders of magnitude. Furthermore, the ML-HK errors in density (as evaluated on a grid in the molecular plane of benzene) are also considerably smaller than the difference in density between density functionals (PBE versus LDA\cite{PZLDA}). This result verifies that the ML-HK map is specific to the density used to train the model and should be able to differentiate between densities generated with other electronic structure approaches.

Ethane ($\mathsf{C_2H_6}$), with a small energy barrier for the relative rotation of the methyl groups, is also evaluated in the same way. Using geometries sampled using K-means from 300~K and 350~K classical trajectories, the ML-HK model reproduces the energy of conformers with a MAE of 0.23~kcal/mol for an independent MD trajectory at 300~K (Fig.~\ref{fig:trajectories}b). This test set includes conformers from the sparsely-sampled eclipsed configuration (see supplement Fig.~\ref{sup-fig:ethane-dataset}). Using points from a 400~K trajectory improves the ML-HK map due to the increased probability of higher energy rotamers in the training set (see Table~\ref{tab:results_mol}). The training set could also be constructed by including explicit rotational conformers, as is common for fitting classical force field parameters\cite{GAFF}. In either case, generating appropriate conformers for training via computationally cheap classical MD significantly decreases the cost of the ML-HK approach.

\begin{figure*}[htb]
  \includegraphics[width=\textwidth]{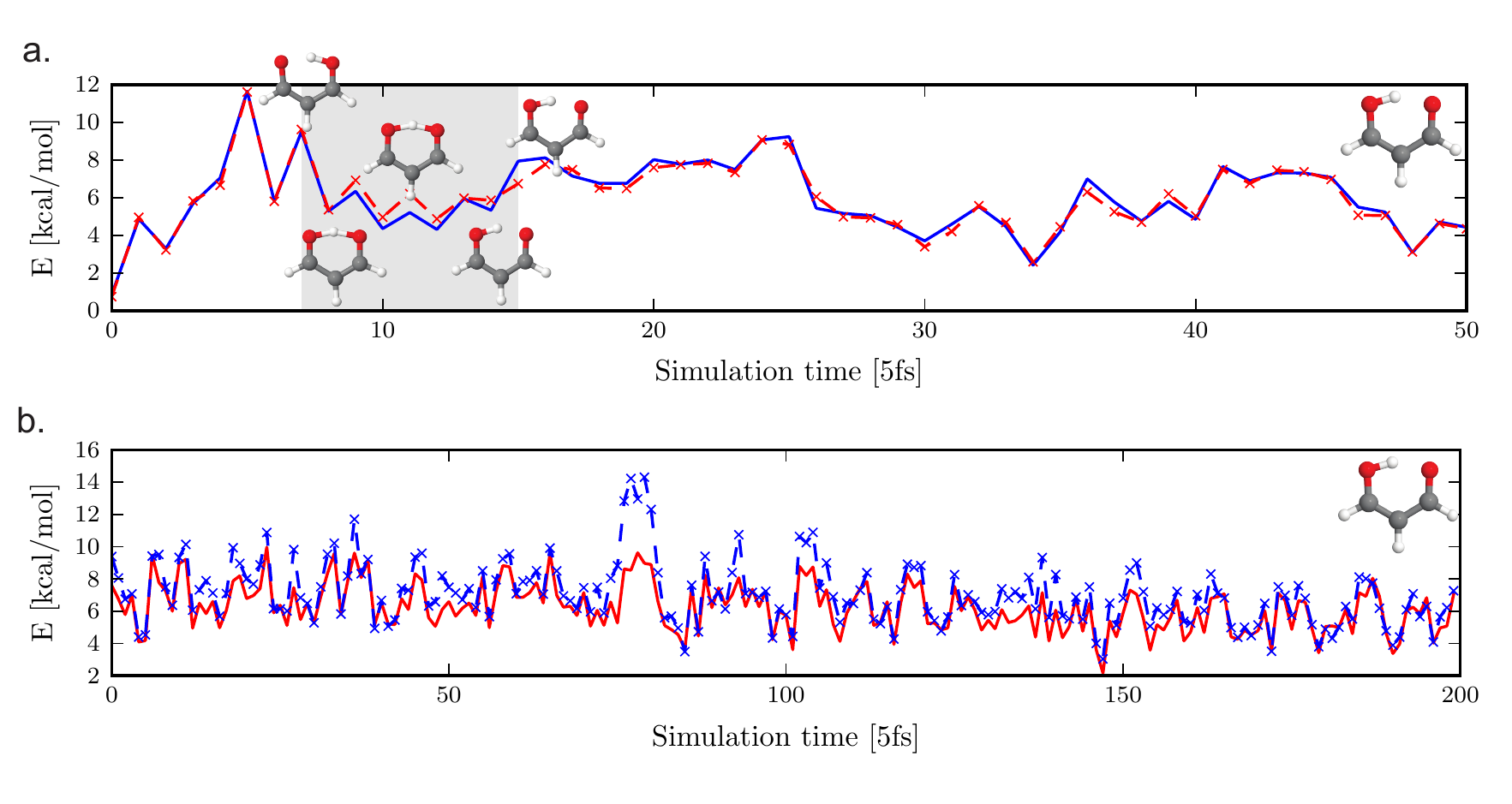}
  \caption{\label{fig:trajectories_malonaldehyde}
  \textbf{a.} Energy errors of ML-HK along a 0.25 ps \textit{ab initio} MD trajectory of malonaldehyde. PBE values in blue, ML-HK values in red. The ML model correctly predicts energies during a proton transfer in frames 7 to 15 without explicitly including these geometries in the training set.
  \textbf{b.} Energy errors of ML-HK along a 1 ps MD trajectory of malonaldehyde \textit{generated by the ML-HK model}. ML-HK values in red, PBE values of trajectory snapshots in blue.}
\end{figure*}

As additional proof of the versatility of the ML-HK map, we show that this approach is also able to interpolate energies for proton transfer in the enol form of malonaldehyde ($\mathsf{C_3H_4O_2}$).
This molecule is a well-known example of intramolecular proton transfer, and our previous AIMD and ab initio path integral studies~\cite{MT_PRL} found classical and quantum free energy barrier values of 3.5 and 1.6 kcal/mol, respectively, from gradient-corrected DFT.
In this work, classical MD trajectories are run for each tautomer separately, with a fixed bonding scheme, then combined for K-means sampling to create the grand training set. The training set also includes an artificially constructed geometry that is the average of tautomer atomic positions. For the test set, we use snapshots from a computationally expensive Born-Oppenheimer ab initio MD trajectory at 300~K. Fig.~\ref{fig:trajectories_malonaldehyde}a shows that the ML-HK map is able to predict DFT energies during a proton transfer event (MAE of 0.27~kcal/mol) despite being trained on classical geometries that did not include these intermediate points.

\begin{figure}[h]
\includegraphics{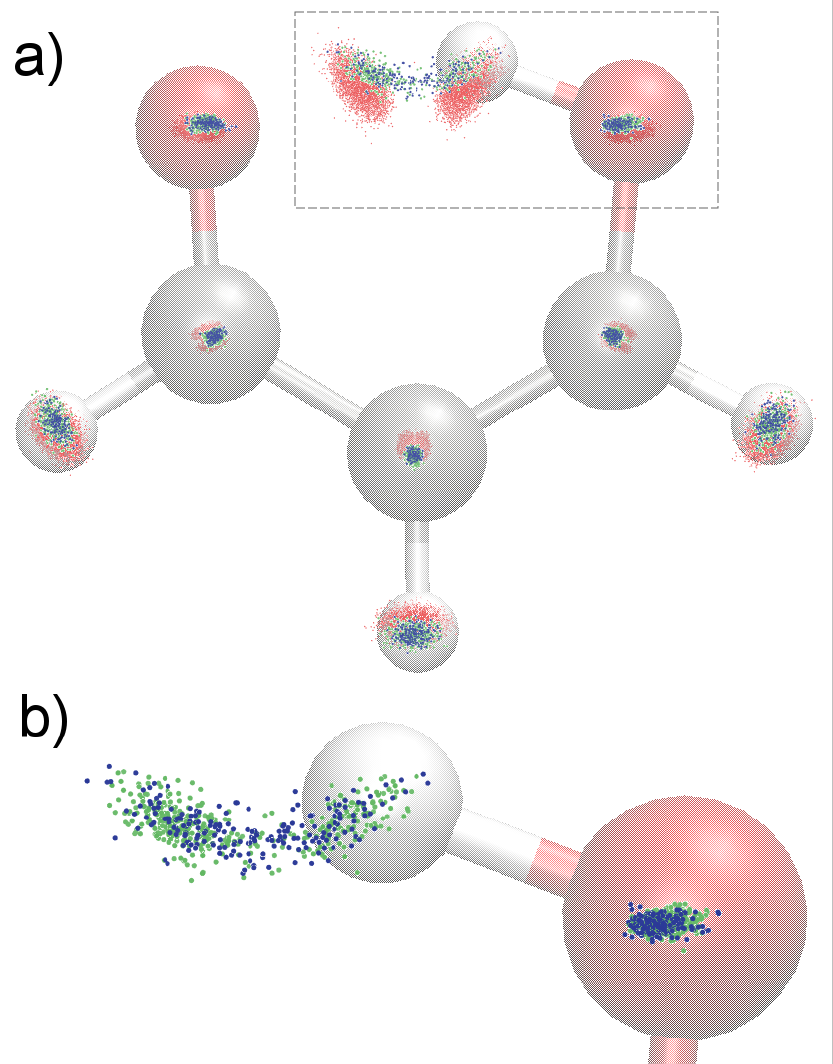}
\caption{\label{fig:malon-dataset} The extent of the malonaldehyde conformers generated by all MD methods. a) The training set of 2,000 representative conformers selected from the classical MD trajectories (red points) by K-means sampling. Test points from an ab initio MD trajectory (green) and the independently generated MD trajectory using the ML-HK model (blue) sample the same coordinate space (offset from the molecular plane for clarity). b) A closer view of the region outlined with a dashed box for the ab initio (green) and ML-HK (blue) trajectories.}
\end{figure}

The ML-HK map can also be used to \textit{generate} a stable MD trajectory for malonaldehyde at 300~K (see Fig.~\ref{fig:trajectories_malonaldehyde}b). In principle, analytic gradients could be obtained for each timestep, but for this first proof-of-concept trajectory, a finite-difference approach was used to determine atomic forces. The ML-HK-generated trajectory samples the same molecular configurations as the ab inito simulation (see Fig.~\ref{fig:malon-dataset}), with mean absolute energy errors of 0.77~kcal/mol, but it typically underestimates the energy for out-of-plane molecular fluctuations at the extremes of the classical training set (maximum error of 5.7~kcal/mol). Even with underestimated energy values, the atomic forces are sufficiently large to return the molecule to the equilibrium configuration, resulting in a stable and long trajectory. The new set of coordinates could be further sampled to expand the training set in a self-consistent manner.  Using iterative ML-HK-generated MD trajectories would eliminate the need to run computationally expensive MD simulations with DFT and would provide an iterative approach to reduce the energy errors for conformations not included in the classical training set.

\begin{table}
  \begin{ruledtabular}
    \begin{tabular}{cc|cccc}
    & \multirow{2}{*}[-0.6em]{\parbox{1.8cm}{Training trajectories}} & \multicolumn{2}{c}{$\Delta E$} & \multicolumn{2}{c}{$\Delta E_D^{\mathrm{ML}}$}\\
    \noalign{\smallskip}
    \cline{3-4}\cline{5-6}
    \noalign{\smallskip}
    Molecule & & MAE & max & MAE & max\\
    \noalign{\smallskip}
    \colrule
    \noalign{\smallskip}
     & 300K & 0.42 & 1.7 & 0.32 & 1.5\\
    Benzene & 300K + 350K & 0.37 & 1.8 & 0.28 & 1.5\\
     & 300K + 400K & 0.47 & 2.3 & 0.30 & 1.8\\
    \noalign{\smallskip}
    \colrule
    \noalign{\smallskip}
     & 300K & 0.20 & 1.5 & 0.17 & 1.3\\
    Ethane & 300K + 350K & 0.23 & 1.4 & 0.19 & 1.1\\
     & 300K + 400K & 0.14 & 1.7 & 0.098 & 0.62\\
    \noalign{\smallskip}
    \colrule
    \noalign{\smallskip}
    Malonaldehyde & 300K + 350K & 0.27 & 1.2 & 0.21 & 0.74\\
    \end{tabular}
\end{ruledtabular}
\caption{\label{tab:results_mol}
Energy and density-driven errors (kcal/mol) of the ML-HK approach on the MD datasets for different training trajectory combinations.
}
\end{table}

\section{Discussion}

For several decades, density functional theory has been a cross-disciplinary area between
theoretical physics, chemistry, and materials sciences.  The methods
of each field cross-fertilize  advances in
other fields.  This has led to its enormous popularity and widespread success, despite
its well-known limitations in both accuracy and the systems and properties to which
it can be applied.

The present work makes a key step forward toward adding an entirely new ingredient to this
mix, namely the construction of functionals via machine learning.  While previous work
showed proofs of principle in 1-D, this is the first demonstration in 3-D, using real molecules
and production-level codes.
We also demonstrate that molecular conformers used in the training set can
be generated by a range of methods, including informed scans and classical MD simulations.
This opens the possibility that machine-learning methods,
which complement all existing approaches to functional approximation, could become a new
and very different approach to this problem, with the potential to greatly reduce the
computational cost of routine DFT calculations.

Our new method, directly learning the Hohenberg-Kohn density-potential
map, overcomes a key bottleneck in previous methodologies that arises in 3-D.
Our approach avoids solving an intermediate more
general problem (the gradient descent) to find the solution of
the more specific problem (finding the ground-state density).
This is called transductive inference by the machine learning community
and is thought to be key to
successful statistical inference methods\cite{vapnik2000nature}.
Following a direct prediction approach with the ML-HK map
increases the accuracy consistently on both 1-D examples and 3-D molecules.
We are also able to learn density models that outperform energy models trained on
much more data. This quantitative observation allows us to
conclude that learning density models is much easier than
learning energy models.   Such a finding should be no surprise to practitioners
of the art of functional construction (see, e.g., \citep{RLCE14}), but
the present work quantifies this observation using standard statistical methods.
As the ML-HK map accurately reflects the training densities, more exact methods could also be used to generate the training set densities for functional development.

We have also derived a way to use basis functions to make the approach
computationally feasible. This makes it easier to integrate the method into existing
DFT codes. Another advantage is the possibility to take the
innate structure of the densities into account, i.e.~spatial correlations are
preserved by using low frequency basis functions.  Again, this fits with the intuition
of experienced practitioners in this field, but here we have quantified this in terms of
machine-learned functionals.

Direct prediction of energies (e.g.,~the ML-KS map) always has
the potential to lead to conceptually easier methods.
But such methods must also abandon the insights and effects that have made DFT
a practical and usefully accurate tool over the past half century.
Many usefully accurate DFT approximations already exist, and the corrections
to such approximations can be machine-learned in precisely the same way
as the entire functional has been approximated here\citep{li2016}.  If machine-learning
corrections require less data, the method becomes more powerful by taking advantage
of existing successes.
Furthermore, existing theorems, such as the viral theorem\citep{CLEB11}, might also be
used to directly construct the kinetic energy functional from an ML-HK map.
In the case of orbital-dependent functionals, such as meta-GGA's or global
hybrids, the method presented here must be extended to learn, e.g., the full density
matrix instead of just the density.

We also note that, for all the 3-D calculations shown here, we machine-learned
$E[\n]$, the entire energy (not just the kinetic energy), which includes some
density-functional approximation for XC\@. But, with a quantum chemical code, we
could have trained on much more accurate quantum chemical densities and energies.
Thus, the ML-HK maps in principle allow the construction of (nearly) exact density functionals for molecular systems, with the potential to significantly reduce the computational cost of quantum chemistry based MD simulations.
All this provides useful directions in which
to expand on the results shown here.

\section{Methods}

\textbf{Kohn-Sham Density Functional Theory (KS-DFT)} is a computational electronic structure method that determines the properties of many-body systems by using functionals of the electron density. The foundation is the Hohenberg-Kohn theorem\citep{HK64} that establishes a one-to-one relationship between potential and density, i.e.~at most one potential can give rise to a given ground-state density.

Kohn-Sham DFT avoids direct approximation of many body effects by
imagining a fictitious system of non-interacting electrons with
the same density as the real one\citep{KS65}.
Its accuracy is limited by the accuracy of existing approximations to
the unknown exchange-correlation energy, while
its computational bottleneck is the solution of the
Kohn-Sham equations that describe the non-interacting particles.

Here, 3-D DFT calculations for ML models are performed with the
Quantum ESPRESSO code\cite{giannozzi2009quantum} using the PBE
exchange-correlation functional\cite{perdew1996generalized} and projector
augmented waves (PAWs)\cite{kresse1999ultrasoft,blochl1994projector} with
Troullier-Martin pseudization for describing the ionic cores\cite{TM91}.
All molecules are simulated in a cubic box (L~=~20~bohr) with a wave function cutoff of 90~Ry. The 1-D dataset is taken from \citet{snyder2012}.

\textbf{Kernel Ridge Regression (KRR)}\citep{hastie2009elements,vu2015understanding} is a machine learning method for regression. It is a kernelized version of Ridge Regression which minimizes the least squares error and applies an $\ell_2$ (Tikhonov) regularization. Let $x_1, \dots, x_m \in \mathbb{R}^d$ be the training data points and let $\mathbf{Y} = {\left(y_1, \dots, y_m \right)}^T$ be their respective labels. KRR then optimizes

\begin{align}
  \min_{\alpha} \sum_{i=1}^m \left|y_i - \sum_{j=1}^m \alpha_j k(x_i, x_j)\right|^2 + \lambda \boldsymbol{\alpha}^\intercal \mathbf{K}\boldsymbol{\alpha}
\end{align}
where $k$ is the kernel function and $\lambda$ is a regularization parameter. $\mathbf{K}$ is the kernel matrix with $\mathbf{K}_{ij} = k(x_i, x_j)$. It admits an analytical solution

\begin{align}
    \boldsymbol{\alpha} = {\left(\mathbf{K} + \lambda \mathbf{I}\right)}^{-1} \mathbf{Y}.
\end{align}
Most popular is the Gaussian (radial basis function) kernel which allows to find a smooth non-linear model function in input space that corresponds to a linear function in an infinite dimensional feature space\citep{muller2001introduction}.

For the ML-HK map, the canonical error is given by the $\mathcal{L}_2$ distance between predicted and true densities

\begin{align}
e(\boldsymbol{\beta}) &= \sum_{i=1}^M \lVert n_i - n^\text{ML}[v_i] \rVert_{\mathcal{L}_2} \\
          &= \sum_{i=1}^M \left\lVert n_i - \sum_{l=1}^L \sum_{j=1}^M \beta_j^{(l)} k(v_i, v_j) \phi_l \right\rVert_{\mathcal{L}_2}.
\end{align}
The ML model coefficients $\boldsymbol{\beta}^{(l)}$ can be optimized independently for each basis coefficient $l$ via

\begin{align}
\boldsymbol{\beta}^{(l)} = {\left(\mathbf{K}_{\sigma^{(l)}} + \lambda^{(l)} \mathbf{I}\right)}^{-1} \mathbf{u}^{(l)}, \quad l = 1, \dots, L.
\end{align}

\textbf{Cross-validation.} Note that all model parameters and hyper-parameters are estimated on the training set; the  hyper-parameter choice makes use of standard cross-validation procedures (see \citet{hansen2013}). Once the model is fixed after training, it is applied unchanged out-of-sample.

\textbf{Exact calculations.} Relative energy errors of the ML models trained on KS-DFT calculations are determined by comparing to accurate energies from the Molpro Quantum Chemistry Software\cite{MOLPRO_brief} using the Full Configuration Interaction method for $\mathsf{H_2}$ and CCSD(T)\cite{AKW07} for $\mathsf{H_2O}$.

\textbf{Molecular Dynamics (MD).} For benzene, ethane, and malonaldehyde, GAFF parameters\cite{GAFF} were assigned using the AmberTools package\cite{antechamber}. Geometry optimizations were performed using MP2/6-31g(d) in Gaussian09\cite{Gaussian09}. Atomic charge assignments are from RESP fits to HF/6-31g(d) calculations at optimized geometries\cite{RESP} and two additional rotational conformers for ethane.

For the three larger molecules, classical isothermal MD simulations were run using the PINY\_MD package\cite{PINYMD} with massive Nos\'{e}-Hoover chain (NHC) thermostats\cite{nhc} for atomic degrees of freedom (length~=~4, $\tau$~=~20~fs, Suzuki-Yoshida order~=~7, multiple time step~=~4) and a time step of 1~fs. The r-RESPA multiple time step approach\cite{respa} was employed to compute rapidly varying forces more frequently (torsions every 0.5~fs, bonds/bends every 0.1~fs). Systems were equilibrated for 100~ps before collecting snapshots every 100~fs from 1~ns trajectories. Snapshots were aligned to a reference molecule prior to DFT calculations for the ML model. For malonaldehyde, the ML training set geometries were selected from trajectories for both enol tautomers as the GAFF force field does not permit changes in chemical bond environments.

For malonaldehyde, an additional Born-Oppenheimer MD simulation using DFT was run using the QUICKSTEP package\cite{QS} in CP2K v. 2.6.2\cite{CP2K}. The PBE exchange-correlation functional\cite{perdew1996generalized} was used in the Gaussian and plane wave (GPW) scheme\cite{GPW} with DZVP-MOLOPT-GTH (m-DZVP) basis sets\cite{MOLOPT} paired with the appropriate dual-space GTH pseudopotentials\cite{GTHpp} optimized for the PBE functional\cite{PBEpp}.  Wave functions were converged to 1E-7~Hartree using the orbital transformation method\cite{OT} on a multiple grid (n~=~5) with a cutoff of 900~Ry for the system in a cubic box (L~=~20~bohr). A temperature of 300~K was maintained using massive NHC thermostats\cite{nhc} (length~=~4, $\tau$~=~10~fs, Suzuki-Yoshida order~=~7, multiple time step~=~4) and a time step of 0.5~fs.

In order to generate the MD trajectory with the ML-HK model, we used the Atomistic Simulation Environment (ASE) \cite{BJ02} with a 0.5~fs timestep and a temperature of 300~K maintained via a Langevin thermostat. A thermostat friction value of 0.01~atomic units (0.413~fs$^{-1}$) was chosen to reproduce the fluctuations in C atoms observed for the DFT-based trajectory (see ESI). In this proof-of-concept work, atomic forces were calculated using central finite differences, with $\epsilon$~=~0.001~\AA\ chosen to conserve the total energy during the simulation. The last 1~ps of a 4~ps trajectory was used to evaluate the performance of the ML-HK model.

\section{Data availability}
All datasets used in this work are available at \url{http://quantum- machine.org/datasets/}.

\section{Acknowledgments}
We thank J.C.~Snyder for early discussions and H.~Glawe for helpful guidance regarding the 3-D reference computations. We thank IPAM at UCLA for repeated hospitality. Work at UC Irvine supported by NSF CHE-1464795. KRM and FB thank the Einstein Foundation for generously funding the ETERNAL project. This work was supported by Institute for Information \& Communications Technology Promotion (IITP) grant funded by the Korea government (No.\ 2017-0-00451). Work at NYU supported by the U.S.\ Army Research Office under contract/grant number W911NF-13-1-0387 (MET and LV). Ab initio trajectory was run using High Performance Computing resources at NYU\@. Other DFT simulations were run using High Performance Computing resources at MPI Halle.

\section{Author contributions}
FB performed DFT calculations and ML experiments. LV performed classical and ab initio MD simulations. LL performed FCI and CCSD(T) calculations. KB and KRM initiated the work and contributed to the theory and experiments. All authors contributed to the manuscript.

\bibliography{bypassing-ks-with-ml.bib}

\begin{thebibliography}{65}%
\makeatletter
\providecommand \@ifxundefined [1]{%
 \@ifx{#1\undefined}
}%
\providecommand \@ifnum [1]{%
 \ifnum #1\expandafter \@firstoftwo
 \else \expandafter \@secondoftwo
 \fi
}%
\providecommand \@ifx [1]{%
 \ifx #1\expandafter \@firstoftwo
 \else \expandafter \@secondoftwo
 \fi
}%
\providecommand \natexlab [1]{#1}%
\providecommand \enquote  [1]{``#1''}%
\providecommand \bibnamefont  [1]{#1}%
\providecommand \bibfnamefont [1]{#1}%
\providecommand \citenamefont [1]{#1}%
\providecommand \href@noop [0]{\@secondoftwo}%
\providecommand \href [0]{\begingroup \@sanitize@url \@href}%
\providecommand \@href[1]{\@@startlink{#1}\@@href}%
\providecommand \@@href[1]{\endgroup#1\@@endlink}%
\providecommand \@sanitize@url [0]{\catcode `\\12\catcode `\$12\catcode
  `\&12\catcode `\#12\catcode `\^12\catcode `\_12\catcode `\%12\relax}%
\providecommand \@@startlink[1]{}%
\providecommand \@@endlink[0]{}%
\providecommand \url  [0]{\begingroup\@sanitize@url \@url }%
\providecommand \@url [1]{\endgroup\@href {#1}{\urlprefix }}%
\providecommand \urlprefix  [0]{URL }%
\providecommand \Eprint [0]{\href }%
\providecommand \doibase [0]{http://dx.doi.org/}%
\providecommand \selectlanguage [0]{\@gobble}%
\providecommand \bibinfo  [0]{\@secondoftwo}%
\providecommand \bibfield  [0]{\@secondoftwo}%
\providecommand \translation [1]{[#1]}%
\providecommand \BibitemOpen [0]{}%
\providecommand \bibitemStop [0]{}%
\providecommand \bibitemNoStop [0]{.\EOS\space}%
\providecommand \EOS [0]{\spacefactor3000\relax}%
\providecommand \BibitemShut  [1]{\csname bibitem#1\endcsname}%
\let\auto@bib@innerbib\@empty
\bibitem [{\citenamefont {Kohn}\ and\ \citenamefont {Sham}(1965)}]{KS65}%
  \BibitemOpen
  \bibfield  {author} {\bibinfo {author} {\bibfnamefont {W.}~\bibnamefont
  {Kohn}}\ and\ \bibinfo {author} {\bibfnamefont {L.~J.}\ \bibnamefont
  {Sham}},\ }\href {\doibase 10.1103/PhysRev.140.A1133} {\bibfield  {journal}
  {\bibinfo  {journal} {Phys. Rev.}\ }\textbf {\bibinfo {volume} {140}},\
  \bibinfo {pages} {A1133} (\bibinfo {year} {1965})}\BibitemShut {NoStop}%
\bibitem [{\citenamefont {Pribram-Jones}\ \emph {et~al.}(2015)\citenamefont
  {Pribram-Jones}, \citenamefont {Gross},\ and\ \citenamefont {Burke}}]{PGB15}%
  \BibitemOpen
  \bibfield  {author} {\bibinfo {author} {\bibfnamefont {A.}~\bibnamefont
  {Pribram-Jones}}, \bibinfo {author} {\bibfnamefont {D.~A.}\ \bibnamefont
  {Gross}}, \ and\ \bibinfo {author} {\bibfnamefont {K.}~\bibnamefont
  {Burke}},\ }\href {\doibase 10.1146/annurev-physchem-040214-121420}
  {\bibfield  {journal} {\bibinfo  {journal} {Annual Review of Physical
  Chemistry}\ }\textbf {\bibinfo {volume} {66}},\ \bibinfo {pages} {283}
  (\bibinfo {year} {2015})}\BibitemShut {NoStop}%
\bibitem [{\citenamefont {Perdew}(1986)}]{P86}%
  \BibitemOpen
  \bibfield  {author} {\bibinfo {author} {\bibfnamefont {J.~P.}\ \bibnamefont
  {Perdew}},\ }\href {\doibase 10.1103/PhysRevB.33.8822} {\bibfield  {journal}
  {\bibinfo  {journal} {Phys. Rev. B}\ }\textbf {\bibinfo {volume} {33}},\
  \bibinfo {pages} {8822} (\bibinfo {year} {1986})}\BibitemShut {NoStop}%
\bibitem [{\citenamefont {Becke}(1993)}]{B93}%
  \BibitemOpen
  \bibfield  {author} {\bibinfo {author} {\bibfnamefont {A.~D.}\ \bibnamefont
  {Becke}},\ }\href {\doibase http://dx.doi.org/10.1063/1.464913} {\bibfield
  {journal} {\bibinfo  {journal} {The Journal of Chemical Physics}\ }\textbf
  {\bibinfo {volume} {98}},\ \bibinfo {pages} {5648} (\bibinfo {year}
  {1993})}\BibitemShut {NoStop}%
\bibitem [{MGI(2011)}]{MGI}%
  \BibitemOpen
  \href@noop {} {\enquote {\bibinfo {title} {Materials genome initiative for
  global competitiveness},}\ } (\bibinfo {year} {2011})\BibitemShut {NoStop}%
\bibitem [{\citenamefont {Jain}\ \emph {et~al.}(2011)\citenamefont {Jain},
  \citenamefont {Hautier}, \citenamefont {Moore}, \citenamefont {Ong},
  \citenamefont {Fischer}, \citenamefont {Mueller}, \citenamefont {Persson},\
  and\ \citenamefont {Ceder}}]{jain2011}%
  \BibitemOpen
  \bibfield  {author} {\bibinfo {author} {\bibfnamefont {A.}~\bibnamefont
  {Jain}}, \bibinfo {author} {\bibfnamefont {G.}~\bibnamefont {Hautier}},
  \bibinfo {author} {\bibfnamefont {C.~J.}\ \bibnamefont {Moore}}, \bibinfo
  {author} {\bibfnamefont {S.~P.}\ \bibnamefont {Ong}}, \bibinfo {author}
  {\bibfnamefont {C.~C.}\ \bibnamefont {Fischer}}, \bibinfo {author}
  {\bibfnamefont {T.}~\bibnamefont {Mueller}}, \bibinfo {author} {\bibfnamefont
  {K.~A.}\ \bibnamefont {Persson}}, \ and\ \bibinfo {author} {\bibfnamefont
  {G.}~\bibnamefont {Ceder}},\ }\href {\doibase
  http://dx.doi.org/10.1016/j.commatsci.2011.02.023} {\bibfield  {journal}
  {\bibinfo  {journal} {Computational Materials Science}\ }\textbf {\bibinfo
  {volume} {50}},\ \bibinfo {pages} {2295 } (\bibinfo {year}
  {2011})}\BibitemShut {NoStop}%
\bibitem [{\citenamefont {Pozun}\ \emph {et~al.}(2012)\citenamefont {Pozun},
  \citenamefont {Hansen}, \citenamefont {Sheppard}, \citenamefont {Rupp},
  \citenamefont {M\"{u}ller},\ and\ \citenamefont {Henkelman}}]{PHSR12}%
  \BibitemOpen
  \bibfield  {author} {\bibinfo {author} {\bibfnamefont {Z.~D.}\ \bibnamefont
  {Pozun}}, \bibinfo {author} {\bibfnamefont {K.}~\bibnamefont {Hansen}},
  \bibinfo {author} {\bibfnamefont {D.}~\bibnamefont {Sheppard}}, \bibinfo
  {author} {\bibfnamefont {M.}~\bibnamefont {Rupp}}, \bibinfo {author}
  {\bibfnamefont {K.-R.}\ \bibnamefont {M\"{u}ller}}, \ and\ \bibinfo {author}
  {\bibfnamefont {G.}~\bibnamefont {Henkelman}},\ }\href {\doibase
  10.1063/1.4707167} {\bibfield  {journal} {\bibinfo  {journal} {The Journal of
  Chemical Physics}\ }\textbf {\bibinfo {volume} {136}},\ \bibinfo {eid}
  {174101} (\bibinfo {year} {2012})}\BibitemShut {NoStop}%
\bibitem [{\citenamefont {McGibbon}\ and\ \citenamefont {Pande}(2013)}]{MP13}%
  \BibitemOpen
  \bibfield  {author} {\bibinfo {author} {\bibfnamefont {R.~T.}\ \bibnamefont
  {McGibbon}}\ and\ \bibinfo {author} {\bibfnamefont {V.~S.}\ \bibnamefont
  {Pande}},\ }\href@noop {} {\bibfield  {journal} {\bibinfo  {journal} {Journal
  of Chemical Theory and Computation}\ }\textbf {\bibinfo {volume} {9}},\
  \bibinfo {pages} {2900} (\bibinfo {year} {2013})}\BibitemShut {NoStop}%
\bibitem [{\citenamefont {Fletcher}\ \emph {et~al.}(2014)\citenamefont
  {Fletcher}, \citenamefont {Davie},\ and\ \citenamefont {Popelier}}]{FDP14}%
  \BibitemOpen
  \bibfield  {author} {\bibinfo {author} {\bibfnamefont {T.~L.}\ \bibnamefont
  {Fletcher}}, \bibinfo {author} {\bibfnamefont {S.~J.}\ \bibnamefont {Davie}},
  \ and\ \bibinfo {author} {\bibfnamefont {P.~L.}\ \bibnamefont {Popelier}},\
  }\href@noop {} {\bibfield  {journal} {\bibinfo  {journal} {Journal of
  chemical theory and computation}\ }\textbf {\bibinfo {volume} {10}},\
  \bibinfo {pages} {3708} (\bibinfo {year} {2014})}\BibitemShut {NoStop}%
\bibitem [{\citenamefont {Rupp}\ \emph {et~al.}(2012)\citenamefont {Rupp},
  \citenamefont {Tkatchenko}, \citenamefont {Müller},\ and\ \citenamefont {von
  Lilienfeld}}]{rupp2012fast}%
  \BibitemOpen
  \bibfield  {author} {\bibinfo {author} {\bibfnamefont {M.}~\bibnamefont
  {Rupp}}, \bibinfo {author} {\bibfnamefont {A.}~\bibnamefont {Tkatchenko}},
  \bibinfo {author} {\bibfnamefont {K.-R.}\ \bibnamefont {Müller}}, \ and\
  \bibinfo {author} {\bibfnamefont {O.~A.}\ \bibnamefont {von Lilienfeld}},\
  }\href {\doibase 10.1103/PhysRevLett.108.058301} {\bibfield  {journal}
  {\bibinfo  {journal} {Phys. Rev. Lett.}\ }\textbf {\bibinfo {volume} {108}},\
  \bibinfo {pages} {058301} (\bibinfo {year} {2012})}\BibitemShut {NoStop}%
\bibitem [{\citenamefont {Hautier}\ \emph {et~al.}(2010)\citenamefont
  {Hautier}, \citenamefont {Fischer}, \citenamefont {Jain}, \citenamefont
  {Mueller},\ and\ \citenamefont {Ceder}}]{hautier2010finding}%
  \BibitemOpen
  \bibfield  {author} {\bibinfo {author} {\bibfnamefont {G.}~\bibnamefont
  {Hautier}}, \bibinfo {author} {\bibfnamefont {C.~C.}\ \bibnamefont
  {Fischer}}, \bibinfo {author} {\bibfnamefont {A.}~\bibnamefont {Jain}},
  \bibinfo {author} {\bibfnamefont {T.}~\bibnamefont {Mueller}}, \ and\
  \bibinfo {author} {\bibfnamefont {G.}~\bibnamefont {Ceder}},\ }\href
  {\doibase 10.1021/cm100795d} {\bibfield  {journal} {\bibinfo  {journal}
  {Chem. Mater.}\ }\textbf {\bibinfo {volume} {22}},\ \bibinfo {pages} {3762}
  (\bibinfo {year} {2010})}\BibitemShut {NoStop}%
\bibitem [{\citenamefont {Hansen}\ \emph {et~al.}(2013)\citenamefont {Hansen},
  \citenamefont {Montavon}, \citenamefont {Biegler}, \citenamefont {Fazli},
  \citenamefont {Rupp}, \citenamefont {Scheffler}, \citenamefont {von
  Lilienfeld}, \citenamefont {Tkatchenko},\ and\ \citenamefont
  {Müller}}]{hansen2013}%
  \BibitemOpen
  \bibfield  {author} {\bibinfo {author} {\bibfnamefont {K.}~\bibnamefont
  {Hansen}}, \bibinfo {author} {\bibfnamefont {G.}~\bibnamefont {Montavon}},
  \bibinfo {author} {\bibfnamefont {F.}~\bibnamefont {Biegler}}, \bibinfo
  {author} {\bibfnamefont {S.}~\bibnamefont {Fazli}}, \bibinfo {author}
  {\bibfnamefont {M.}~\bibnamefont {Rupp}}, \bibinfo {author} {\bibfnamefont
  {M.}~\bibnamefont {Scheffler}}, \bibinfo {author} {\bibfnamefont {O.~A.}\
  \bibnamefont {von Lilienfeld}}, \bibinfo {author} {\bibfnamefont
  {A.}~\bibnamefont {Tkatchenko}}, \ and\ \bibinfo {author} {\bibfnamefont
  {K.-R.}\ \bibnamefont {Müller}},\ }\href {\doibase 10.1021/ct400195d}
  {\bibfield  {journal} {\bibinfo  {journal} {J. Chem. Theory Comput.}\
  }\textbf {\bibinfo {volume} {9}},\ \bibinfo {pages} {3404} (\bibinfo {year}
  {2013})}\BibitemShut {NoStop}%
\bibitem [{\citenamefont {Schütt}\ \emph {et~al.}(2014)\citenamefont
  {Schütt}, \citenamefont {Glawe}, \citenamefont {Brockherde}, \citenamefont
  {Sanna}, \citenamefont {Müller},\ and\ \citenamefont {Gross}}]{schutt2014}%
  \BibitemOpen
  \bibfield  {author} {\bibinfo {author} {\bibfnamefont {K.~T.}\ \bibnamefont
  {Schütt}}, \bibinfo {author} {\bibfnamefont {H.}~\bibnamefont {Glawe}},
  \bibinfo {author} {\bibfnamefont {F.}~\bibnamefont {Brockherde}}, \bibinfo
  {author} {\bibfnamefont {A.}~\bibnamefont {Sanna}}, \bibinfo {author}
  {\bibfnamefont {K.-R.}\ \bibnamefont {Müller}}, \ and\ \bibinfo {author}
  {\bibfnamefont {E.~K.~U.}\ \bibnamefont {Gross}},\ }\href {\doibase
  10.1103/PhysRevB.89.205118} {\bibfield  {journal} {\bibinfo  {journal} {Phys.
  Rev. B}\ }\textbf {\bibinfo {volume} {89}},\ \bibinfo {pages} {205118}
  (\bibinfo {year} {2014})}\BibitemShut {NoStop}%
\bibitem [{\citenamefont {Hansen}\ \emph {et~al.}(2015)\citenamefont {Hansen},
  \citenamefont {Biegler}, \citenamefont {Ramakrishnan}, \citenamefont
  {Pronobis}, \citenamefont {von Lilienfeld}, \citenamefont {Müller},\ and\
  \citenamefont {Tkatchenko}}]{HBRP15}%
  \BibitemOpen
  \bibfield  {author} {\bibinfo {author} {\bibfnamefont {K.}~\bibnamefont
  {Hansen}}, \bibinfo {author} {\bibfnamefont {F.}~\bibnamefont {Biegler}},
  \bibinfo {author} {\bibfnamefont {R.}~\bibnamefont {Ramakrishnan}}, \bibinfo
  {author} {\bibfnamefont {W.}~\bibnamefont {Pronobis}}, \bibinfo {author}
  {\bibfnamefont {O.~A.}\ \bibnamefont {von Lilienfeld}}, \bibinfo {author}
  {\bibfnamefont {K.-R.}\ \bibnamefont {Müller}}, \ and\ \bibinfo {author}
  {\bibfnamefont {A.}~\bibnamefont {Tkatchenko}},\ }\href {\doibase
  10.1021/acs.jpclett.5b00831} {\bibfield  {journal} {\bibinfo  {journal} {The
  Journal of Physical Chemistry Letters}\ }\textbf {\bibinfo {volume} {6}},\
  \bibinfo {pages} {2326} (\bibinfo {year} {2015})},\ \bibinfo {note} {pMID:
  26113956}\BibitemShut {NoStop}%
\bibitem [{\citenamefont {Sch{\"{u}}tt}\ \emph {et~al.}(2017)\citenamefont
  {Sch{\"{u}}tt}, \citenamefont {Arbabzadah}, \citenamefont {Chmiela},
  \citenamefont {M{\"{u}}ller},\ and\ \citenamefont {Tkatchenko}}]{Schutt2017}%
  \BibitemOpen
  \bibfield  {author} {\bibinfo {author} {\bibfnamefont {K.~T.}\ \bibnamefont
  {Sch{\"{u}}tt}}, \bibinfo {author} {\bibfnamefont {F.}~\bibnamefont
  {Arbabzadah}}, \bibinfo {author} {\bibfnamefont {S.}~\bibnamefont {Chmiela}},
  \bibinfo {author} {\bibfnamefont {K.~R.}\ \bibnamefont {M{\"{u}}ller}}, \
  and\ \bibinfo {author} {\bibfnamefont {A.}~\bibnamefont {Tkatchenko}},\
  }\href {\doibase 10.1038/ncomms13890} {\bibfield  {journal} {\bibinfo
  {journal} {Nat. Commun.}\ }\textbf {\bibinfo {volume} {8}},\ \bibinfo {pages}
  {13890} (\bibinfo {year} {2017})}\BibitemShut {NoStop}%
\bibitem [{\citenamefont {Behler}\ and\ \citenamefont
  {Parrinello}(2007)}]{BehlerParinello2007}%
  \BibitemOpen
  \bibfield  {author} {\bibinfo {author} {\bibfnamefont {J.}~\bibnamefont
  {Behler}}\ and\ \bibinfo {author} {\bibfnamefont {M.}~\bibnamefont
  {Parrinello}},\ }\href {\doibase 10.1103/PhysRevLett.98.146401} {\bibfield
  {journal} {\bibinfo  {journal} {Phys. Rev. Lett.}\ }\textbf {\bibinfo
  {volume} {98}},\ \bibinfo {pages} {146401} (\bibinfo {year}
  {2007})}\BibitemShut {NoStop}%
\bibitem [{\citenamefont {Seko}\ \emph {et~al.}(2014)\citenamefont {Seko},
  \citenamefont {Takahashi},\ and\ \citenamefont
  {Tanaka}}]{SekoTakahashiTanaka2014}%
  \BibitemOpen
  \bibfield  {author} {\bibinfo {author} {\bibfnamefont {A.}~\bibnamefont
  {Seko}}, \bibinfo {author} {\bibfnamefont {A.}~\bibnamefont {Takahashi}}, \
  and\ \bibinfo {author} {\bibfnamefont {I.}~\bibnamefont {Tanaka}},\ }\href
  {\doibase 10.1103/PhysRevB.90.024101} {\bibfield  {journal} {\bibinfo
  {journal} {Phys. Rev. B}\ }\textbf {\bibinfo {volume} {90}},\ \bibinfo
  {pages} {024101} (\bibinfo {year} {2014})}\BibitemShut {NoStop}%
\bibitem [{\citenamefont {Li}\ \emph {et~al.}(2015)\citenamefont {Li},
  \citenamefont {Kermode},\ and\ \citenamefont {De~Vita}}]{LKD2015}%
  \BibitemOpen
  \bibfield  {author} {\bibinfo {author} {\bibfnamefont {Z.}~\bibnamefont
  {Li}}, \bibinfo {author} {\bibfnamefont {J.~R.}\ \bibnamefont {Kermode}}, \
  and\ \bibinfo {author} {\bibfnamefont {A.}~\bibnamefont {De~Vita}},\ }\href
  {\doibase 10.1103/PhysRevLett.114.096405} {\bibfield  {journal} {\bibinfo
  {journal} {Phys. Rev. Lett.}\ }\textbf {\bibinfo {volume} {114}},\ \bibinfo
  {pages} {096405} (\bibinfo {year} {2015})}\BibitemShut {NoStop}%
\bibitem [{\citenamefont {Chmiela}\ \emph {et~al.}(2017)\citenamefont
  {Chmiela}, \citenamefont {Tkatchenko}, \citenamefont {Sauceda}, \citenamefont
  {Poltavsky}, \citenamefont {Sch{\"{u}}tt},\ and\ \citenamefont
  {M{\"{u}}ller}}]{Chmiela2017}%
  \BibitemOpen
  \bibfield  {author} {\bibinfo {author} {\bibfnamefont {S.}~\bibnamefont
  {Chmiela}}, \bibinfo {author} {\bibfnamefont {A.}~\bibnamefont {Tkatchenko}},
  \bibinfo {author} {\bibfnamefont {H.~E.}\ \bibnamefont {Sauceda}}, \bibinfo
  {author} {\bibfnamefont {I.}~\bibnamefont {Poltavsky}}, \bibinfo {author}
  {\bibfnamefont {K.~T.}\ \bibnamefont {Sch{\"{u}}tt}}, \ and\ \bibinfo
  {author} {\bibfnamefont {K.-R.}\ \bibnamefont {M{\"{u}}ller}},\ }\href
  {\doibase 10.1126/sciadv.1603015} {\bibfield  {journal} {\bibinfo  {journal}
  {Sci. Adv.}\ }\textbf {\bibinfo {volume} {3}},\ \bibinfo {pages} {e1603015}
  (\bibinfo {year} {2017})}\BibitemShut {NoStop}%
\bibitem [{\citenamefont {Snyder}\ \emph {et~al.}(2012)\citenamefont {Snyder},
  \citenamefont {Rupp}, \citenamefont {Hansen}, \citenamefont {Müller},\ and\
  \citenamefont {Burke}}]{snyder2012}%
  \BibitemOpen
  \bibfield  {author} {\bibinfo {author} {\bibfnamefont {J.~C.}\ \bibnamefont
  {Snyder}}, \bibinfo {author} {\bibfnamefont {M.}~\bibnamefont {Rupp}},
  \bibinfo {author} {\bibfnamefont {K.}~\bibnamefont {Hansen}}, \bibinfo
  {author} {\bibfnamefont {K.-R.}\ \bibnamefont {Müller}}, \ and\ \bibinfo
  {author} {\bibfnamefont {K.}~\bibnamefont {Burke}},\ }\href {\doibase
  10.1103/PhysRevLett.108.253002} {\bibfield  {journal} {\bibinfo  {journal}
  {Phys. Rev. Lett.}\ }\textbf {\bibinfo {volume} {108}},\ \bibinfo {pages}
  {253002} (\bibinfo {year} {2012})}\BibitemShut {NoStop}%
\bibitem [{\citenamefont {Snyder}\ \emph
  {et~al.}(2013{\natexlab{a}})\citenamefont {Snyder}, \citenamefont {Rupp},
  \citenamefont {Hansen}, \citenamefont {Blooston}, \citenamefont {Müller},\
  and\ \citenamefont {Burke}}]{snyder2013orbitalfree}%
  \BibitemOpen
  \bibfield  {author} {\bibinfo {author} {\bibfnamefont {J.~C.}\ \bibnamefont
  {Snyder}}, \bibinfo {author} {\bibfnamefont {M.}~\bibnamefont {Rupp}},
  \bibinfo {author} {\bibfnamefont {K.}~\bibnamefont {Hansen}}, \bibinfo
  {author} {\bibfnamefont {L.}~\bibnamefont {Blooston}}, \bibinfo {author}
  {\bibfnamefont {K.-R.}\ \bibnamefont {Müller}}, \ and\ \bibinfo {author}
  {\bibfnamefont {K.}~\bibnamefont {Burke}},\ }\href {\doibase
  10.1063/1.4834075} {\bibfield  {journal} {\bibinfo  {journal} {J. Chem.
  Phys}\ }\textbf {\bibinfo {volume} {139}},\ \bibinfo {pages} {224104}
  (\bibinfo {year} {2013}{\natexlab{a}})}\BibitemShut {NoStop}%
\bibitem [{\citenamefont {Li}\ \emph {et~al.}(2016{\natexlab{a}})\citenamefont
  {Li}, \citenamefont {Snyder}, \citenamefont {Pelaschier}, \citenamefont
  {Huang}, \citenamefont {Niranjan}, \citenamefont {Duncan}, \citenamefont
  {Rupp}, \citenamefont {Müller},\ and\ \citenamefont {Burke}}]{li2015}%
  \BibitemOpen
  \bibfield  {author} {\bibinfo {author} {\bibfnamefont {L.}~\bibnamefont
  {Li}}, \bibinfo {author} {\bibfnamefont {J.~C.}\ \bibnamefont {Snyder}},
  \bibinfo {author} {\bibfnamefont {I.~M.}\ \bibnamefont {Pelaschier}},
  \bibinfo {author} {\bibfnamefont {J.}~\bibnamefont {Huang}}, \bibinfo
  {author} {\bibfnamefont {U.-N.}\ \bibnamefont {Niranjan}}, \bibinfo {author}
  {\bibfnamefont {P.}~\bibnamefont {Duncan}}, \bibinfo {author} {\bibfnamefont
  {M.}~\bibnamefont {Rupp}}, \bibinfo {author} {\bibfnamefont {K.-R.}\
  \bibnamefont {Müller}}, \ and\ \bibinfo {author} {\bibfnamefont
  {K.}~\bibnamefont {Burke}},\ }\href {\doibase 10.1002/qua.25040} {\bibfield
  {journal} {\bibinfo  {journal} {International Journal of Quantum Chemistry}\
  }\textbf {\bibinfo {volume} {116}},\ \bibinfo {pages} {819} (\bibinfo {year}
  {2016}{\natexlab{a}})}\BibitemShut {NoStop}%
\bibitem [{\citenamefont {Li}\ \emph {et~al.}(2016{\natexlab{b}})\citenamefont
  {Li}, \citenamefont {Baker}, \citenamefont {White},\ and\ \citenamefont
  {Burke}}]{li2016}%
  \BibitemOpen
  \bibfield  {author} {\bibinfo {author} {\bibfnamefont {L.}~\bibnamefont
  {Li}}, \bibinfo {author} {\bibfnamefont {T.~E.}\ \bibnamefont {Baker}},
  \bibinfo {author} {\bibfnamefont {S.~R.}\ \bibnamefont {White}}, \ and\
  \bibinfo {author} {\bibfnamefont {K.}~\bibnamefont {Burke}},\ }\href
  {\doibase 10.1103/PhysRevB.94.245129} {\bibfield  {journal} {\bibinfo
  {journal} {Phys. Rev. B}\ }\textbf {\bibinfo {volume} {94}},\ \bibinfo
  {pages} {245129} (\bibinfo {year} {2016}{\natexlab{b}})}\BibitemShut
  {NoStop}%
\bibitem [{\citenamefont {Giannozzi}\ \emph
  {et~al.}(2009{\natexlab{a}})\citenamefont {Giannozzi}, \citenamefont
  {Baroni}, \citenamefont {Bonini}, \citenamefont {Calandra}, \citenamefont
  {Car}, \citenamefont {Cavazzoni}, \citenamefont {Ceresoli}, \citenamefont
  {Chiarotti}, \citenamefont {Cococcioni}, \citenamefont {Dabo}, \citenamefont
  {{Dal Corso}}, \citenamefont {de~Gironcoli}, \citenamefont {Fabris},
  \citenamefont {Fratesi}, \citenamefont {Gebauer}, \citenamefont {Gerstmann},
  \citenamefont {Gougoussis}, \citenamefont {Kokalj}, \citenamefont {Lazzeri},
  \citenamefont {Martin-Samos}, \citenamefont {Marzari}, \citenamefont {Mauri},
  \citenamefont {Mazzarello}, \citenamefont {Paolini}, \citenamefont
  {Pasquarello}, \citenamefont {Paulatto}, \citenamefont {Sbraccia},
  \citenamefont {Scandolo}, \citenamefont {Sclauzero}, \citenamefont
  {Seitsonen}, \citenamefont {Smogunov}, \citenamefont {Umari},\ and\
  \citenamefont {Wentzcovitch}}]{QE2009}%
  \BibitemOpen
  \bibfield  {author} {\bibinfo {author} {\bibfnamefont {P.}~\bibnamefont
  {Giannozzi}}, \bibinfo {author} {\bibfnamefont {S.}~\bibnamefont {Baroni}},
  \bibinfo {author} {\bibfnamefont {N.}~\bibnamefont {Bonini}}, \bibinfo
  {author} {\bibfnamefont {M.}~\bibnamefont {Calandra}}, \bibinfo {author}
  {\bibfnamefont {R.}~\bibnamefont {Car}}, \bibinfo {author} {\bibfnamefont
  {C.}~\bibnamefont {Cavazzoni}}, \bibinfo {author} {\bibfnamefont
  {D.}~\bibnamefont {Ceresoli}}, \bibinfo {author} {\bibfnamefont {G.~L.}\
  \bibnamefont {Chiarotti}}, \bibinfo {author} {\bibfnamefont {M.}~\bibnamefont
  {Cococcioni}}, \bibinfo {author} {\bibfnamefont {I.}~\bibnamefont {Dabo}},
  \bibinfo {author} {\bibfnamefont {A.}~\bibnamefont {{Dal Corso}}}, \bibinfo
  {author} {\bibfnamefont {S.}~\bibnamefont {de~Gironcoli}}, \bibinfo {author}
  {\bibfnamefont {S.}~\bibnamefont {Fabris}}, \bibinfo {author} {\bibfnamefont
  {G.}~\bibnamefont {Fratesi}}, \bibinfo {author} {\bibfnamefont
  {R.}~\bibnamefont {Gebauer}}, \bibinfo {author} {\bibfnamefont
  {U.}~\bibnamefont {Gerstmann}}, \bibinfo {author} {\bibfnamefont
  {C.}~\bibnamefont {Gougoussis}}, \bibinfo {author} {\bibfnamefont
  {A.}~\bibnamefont {Kokalj}}, \bibinfo {author} {\bibfnamefont
  {M.}~\bibnamefont {Lazzeri}}, \bibinfo {author} {\bibfnamefont
  {L.}~\bibnamefont {Martin-Samos}}, \bibinfo {author} {\bibfnamefont
  {N.}~\bibnamefont {Marzari}}, \bibinfo {author} {\bibfnamefont
  {F.}~\bibnamefont {Mauri}}, \bibinfo {author} {\bibfnamefont
  {R.}~\bibnamefont {Mazzarello}}, \bibinfo {author} {\bibfnamefont
  {S.}~\bibnamefont {Paolini}}, \bibinfo {author} {\bibfnamefont
  {A.}~\bibnamefont {Pasquarello}}, \bibinfo {author} {\bibfnamefont
  {L.}~\bibnamefont {Paulatto}}, \bibinfo {author} {\bibfnamefont
  {C.}~\bibnamefont {Sbraccia}}, \bibinfo {author} {\bibfnamefont
  {S.}~\bibnamefont {Scandolo}}, \bibinfo {author} {\bibfnamefont
  {G.}~\bibnamefont {Sclauzero}}, \bibinfo {author} {\bibfnamefont {A.~P.}\
  \bibnamefont {Seitsonen}}, \bibinfo {author} {\bibfnamefont {A.}~\bibnamefont
  {Smogunov}}, \bibinfo {author} {\bibfnamefont {P.}~\bibnamefont {Umari}}, \
  and\ \bibinfo {author} {\bibfnamefont {R.~M.}\ \bibnamefont {Wentzcovitch}},\
  }\href {http://www.quantum-espresso.org} {\bibfield  {journal} {\bibinfo
  {journal} {Journal of Physics: Condensed Matter}\ }\textbf {\bibinfo {volume}
  {21}},\ \bibinfo {pages} {395502 (19pp)} (\bibinfo {year}
  {2009}{\natexlab{a}})}\BibitemShut {NoStop}%
\bibitem [{\citenamefont {Snyder}\ \emph
  {et~al.}(2013{\natexlab{b}})\citenamefont {Snyder}, \citenamefont {Mika},
  \citenamefont {Burke},\ and\ \citenamefont {Müller}}]{snyder2013kernels}%
  \BibitemOpen
  \bibfield  {author} {\bibinfo {author} {\bibfnamefont {J.~C.}\ \bibnamefont
  {Snyder}}, \bibinfo {author} {\bibfnamefont {S.}~\bibnamefont {Mika}},
  \bibinfo {author} {\bibfnamefont {K.}~\bibnamefont {Burke}}, \ and\ \bibinfo
  {author} {\bibfnamefont {K.-R.}\ \bibnamefont {Müller}},\ }in\ \href
  {http://dx.doi.org/10.1007/978-3-642-41136-6_21} {\emph {\bibinfo {booktitle}
  {Empirical Inference}}},\ \bibinfo {editor} {edited by\ \bibinfo {editor}
  {\bibfnamefont {B.}~\bibnamefont {Schölkopf}}, \bibinfo {editor}
  {\bibfnamefont {Z.}~\bibnamefont {Luo}}, \ and\ \bibinfo {editor}
  {\bibfnamefont {V.}~\bibnamefont {Vovk}}}\ (\bibinfo  {publisher} {Springer
  Berlin Heidelberg},\ \bibinfo {year} {2013})\ pp.\ \bibinfo {pages}
  {245--259}\BibitemShut {NoStop}%
\bibitem [{\citenamefont {Snyder}\ \emph {et~al.}(2015)\citenamefont {Snyder},
  \citenamefont {Rupp}, \citenamefont {Müller},\ and\ \citenamefont
  {Burke}}]{snyder2015nonlinear}%
  \BibitemOpen
  \bibfield  {author} {\bibinfo {author} {\bibfnamefont {J.~C.}\ \bibnamefont
  {Snyder}}, \bibinfo {author} {\bibfnamefont {M.}~\bibnamefont {Rupp}},
  \bibinfo {author} {\bibfnamefont {K.-R.}\ \bibnamefont {Müller}}, \ and\
  \bibinfo {author} {\bibfnamefont {K.}~\bibnamefont {Burke}},\ }\href
  {\doibase 10.1002/qua.24937} {\bibfield  {journal} {\bibinfo  {journal} {Int.
  J. Quantum Chem.}\ }\textbf {\bibinfo {volume} {115}},\ \bibinfo {pages}
  {1102} (\bibinfo {year} {2015})}\BibitemShut {NoStop}%
\bibitem [{\citenamefont {Ribeiro}\ \emph {et~al.}(2015)\citenamefont
  {Ribeiro}, \citenamefont {Lee}, \citenamefont {Cangi}, \citenamefont
  {Elliott},\ and\ \citenamefont {Burke}}]{RLCE14}%
  \BibitemOpen
  \bibfield  {author} {\bibinfo {author} {\bibfnamefont {R.~F.}\ \bibnamefont
  {Ribeiro}}, \bibinfo {author} {\bibfnamefont {D.}~\bibnamefont {Lee}},
  \bibinfo {author} {\bibfnamefont {A.}~\bibnamefont {Cangi}}, \bibinfo
  {author} {\bibfnamefont {P.}~\bibnamefont {Elliott}}, \ and\ \bibinfo
  {author} {\bibfnamefont {K.}~\bibnamefont {Burke}},\ }\href {\doibase
  10.1103/PhysRevLett.114.050401} {\bibfield  {journal} {\bibinfo  {journal}
  {Phys. Rev. Lett.}\ }\textbf {\bibinfo {volume} {114}},\ \bibinfo {pages}
  {050401} (\bibinfo {year} {2015})}\BibitemShut {NoStop}%
\bibitem [{\citenamefont {Müller}\ \emph {et~al.}(2001)\citenamefont
  {Müller}, \citenamefont {Mika}, \citenamefont {Rätsch}, \citenamefont
  {Tsuda},\ and\ \citenamefont {Schölkopf}}]{muller2001introduction}%
  \BibitemOpen
  \bibfield  {author} {\bibinfo {author} {\bibfnamefont {K.-R.}\ \bibnamefont
  {Müller}}, \bibinfo {author} {\bibfnamefont {S.}~\bibnamefont {Mika}},
  \bibinfo {author} {\bibfnamefont {G.}~\bibnamefont {Rätsch}}, \bibinfo
  {author} {\bibfnamefont {K.}~\bibnamefont {Tsuda}}, \ and\ \bibinfo {author}
  {\bibfnamefont {B.}~\bibnamefont {Schölkopf}},\ }\href@noop {} {\bibfield
  {journal} {\bibinfo  {journal} {{IEEE} Trans. Neural Netw.}\ }\textbf
  {\bibinfo {volume} {12}},\ \bibinfo {pages} {181} (\bibinfo {year}
  {2001})}\BibitemShut {NoStop}%
\bibitem [{\citenamefont {Kim}\ \emph {et~al.}(2013)\citenamefont {Kim},
  \citenamefont {Sim},\ and\ \citenamefont {Burke}}]{KSB13}%
  \BibitemOpen
  \bibfield  {author} {\bibinfo {author} {\bibfnamefont {M.-C.}\ \bibnamefont
  {Kim}}, \bibinfo {author} {\bibfnamefont {E.}~\bibnamefont {Sim}}, \ and\
  \bibinfo {author} {\bibfnamefont {K.}~\bibnamefont {Burke}},\ }\href
  {\doibase 10.1103/PhysRevLett.111.073003} {\bibfield  {journal} {\bibinfo
  {journal} {Phys. Rev. Lett.}\ }\textbf {\bibinfo {volume} {111}},\ \bibinfo
  {pages} {073003} (\bibinfo {year} {2013})}\BibitemShut {NoStop}%
\bibitem [{\citenamefont {Kim}\ \emph {et~al.}(2014)\citenamefont {Kim},
  \citenamefont {Sim},\ and\ \citenamefont {Burke}}]{KSB14}%
  \BibitemOpen
  \bibfield  {author} {\bibinfo {author} {\bibfnamefont {M.-C.}\ \bibnamefont
  {Kim}}, \bibinfo {author} {\bibfnamefont {E.}~\bibnamefont {Sim}}, \ and\
  \bibinfo {author} {\bibfnamefont {K.}~\bibnamefont {Burke}},\ }\href
  {\doibase http://dx.doi.org/10.1063/1.4869189} {\bibfield  {journal}
  {\bibinfo  {journal} {The Journal of Chemical Physics}\ }\textbf {\bibinfo
  {volume} {140}},\ \bibinfo {pages} {18A528} (\bibinfo {year}
  {2014})}\BibitemShut {NoStop}%
\bibitem [{\citenamefont {Kim}\ \emph {et~al.}(2015)\citenamefont {Kim},
  \citenamefont {Park}, \citenamefont {Son}, \citenamefont {Sim},\ and\
  \citenamefont {Burke}}]{KPSS15}%
  \BibitemOpen
  \bibfield  {author} {\bibinfo {author} {\bibfnamefont {M.-C.}\ \bibnamefont
  {Kim}}, \bibinfo {author} {\bibfnamefont {H.}~\bibnamefont {Park}}, \bibinfo
  {author} {\bibfnamefont {S.}~\bibnamefont {Son}}, \bibinfo {author}
  {\bibfnamefont {E.}~\bibnamefont {Sim}}, \ and\ \bibinfo {author}
  {\bibfnamefont {K.}~\bibnamefont {Burke}},\ }\href {\doibase
  10.1021/acs.jpclett.5b01724} {\bibfield  {journal} {\bibinfo  {journal} {J.
  Phys. Chem. Lett.}\ }\textbf {\bibinfo {volume} {6}},\ \bibinfo {pages}
  {3802} (\bibinfo {year} {2015})}\BibitemShut {NoStop}%
\bibitem [{\citenamefont {Dreizler}\ and\ \citenamefont {Gross}(1990)}]{DG90}%
  \BibitemOpen
  \bibfield  {author} {\bibinfo {author} {\bibfnamefont {R.~M.}\ \bibnamefont
  {Dreizler}}\ and\ \bibinfo {author} {\bibfnamefont {E.~K.~U.}\ \bibnamefont
  {Gross}},\ }\href {\doibase 10.1007/978-3-642-86105-5} {\emph {\bibinfo
  {title} {Density Functional Theory: An Approach to the Quantum Many-Body
  Problem}}}\ (\bibinfo  {publisher} {Springer-Verlag Berlin Heidelberg},\
  \bibinfo {year} {1990})\BibitemShut {NoStop}%
\bibitem [{\citenamefont {Schölkopf}\ \emph {et~al.}(1999)\citenamefont
  {Schölkopf}, \citenamefont {Mika}, \citenamefont {Burges}, \citenamefont
  {Knirsch}, \citenamefont {Müller}, \citenamefont {Rätsch},\ and\
  \citenamefont {Smola}}]{inputspacevs}%
  \BibitemOpen
  \bibfield  {author} {\bibinfo {author} {\bibfnamefont {B.}~\bibnamefont
  {Schölkopf}}, \bibinfo {author} {\bibfnamefont {S.}~\bibnamefont {Mika}},
  \bibinfo {author} {\bibfnamefont {C.}~\bibnamefont {Burges}}, \bibinfo
  {author} {\bibfnamefont {P.}~\bibnamefont {Knirsch}}, \bibinfo {author}
  {\bibfnamefont {K.-R.}\ \bibnamefont {Müller}}, \bibinfo {author}
  {\bibfnamefont {G.}~\bibnamefont {Rätsch}}, \ and\ \bibinfo {author}
  {\bibfnamefont {A.}~\bibnamefont {Smola}},\ }\href {\doibase
  10.1109/72.788641} {\bibfield  {journal} {\bibinfo  {journal} {{IEEE} Trans.
  Neural Netw.}\ }\textbf {\bibinfo {volume} {10}},\ \bibinfo {pages} {1000}
  (\bibinfo {year} {1999})}\BibitemShut {NoStop}%
\bibitem [{\citenamefont {Schölkopf}\ \emph {et~al.}(1998)\citenamefont
  {Schölkopf}, \citenamefont {Smola},\ and\ \citenamefont
  {Müller}}]{scholkopf1998nonlinear}%
  \BibitemOpen
  \bibfield  {author} {\bibinfo {author} {\bibfnamefont {B.}~\bibnamefont
  {Schölkopf}}, \bibinfo {author} {\bibfnamefont {A.}~\bibnamefont {Smola}}, \
  and\ \bibinfo {author} {\bibfnamefont {K.-R.}\ \bibnamefont {Müller}},\
  }\href@noop {} {\bibfield  {journal} {\bibinfo  {journal} {Neural Comput.}\
  }\textbf {\bibinfo {volume} {10}},\ \bibinfo {pages} {1299} (\bibinfo {year}
  {1998})}\BibitemShut {NoStop}%
\bibitem [{\citenamefont {Perdew}\ \emph {et~al.}(1996)\citenamefont {Perdew},
  \citenamefont {Burke},\ and\ \citenamefont
  {Ernzerhof}}]{perdew1996generalized}%
  \BibitemOpen
  \bibfield  {author} {\bibinfo {author} {\bibfnamefont {J.~P.}\ \bibnamefont
  {Perdew}}, \bibinfo {author} {\bibfnamefont {K.}~\bibnamefont {Burke}}, \
  and\ \bibinfo {author} {\bibfnamefont {M.}~\bibnamefont {Ernzerhof}},\ }\href
  {\doibase 10.1103/PhysRevLett.77.3865} {\bibfield  {journal} {\bibinfo
  {journal} {Phys. Rev. Lett.}\ }\textbf {\bibinfo {volume} {77}},\ \bibinfo
  {pages} {3865} (\bibinfo {year} {1996})}\BibitemShut {NoStop}%
\bibitem [{\citenamefont {Kresse}\ and\ \citenamefont
  {Joubert}(1999)}]{kresse1999ultrasoft}%
  \BibitemOpen
  \bibfield  {author} {\bibinfo {author} {\bibfnamefont {G.}~\bibnamefont
  {Kresse}}\ and\ \bibinfo {author} {\bibfnamefont {D.}~\bibnamefont
  {Joubert}},\ }\href {\doibase 10.1103/PhysRevB.59.1758} {\bibfield  {journal}
  {\bibinfo  {journal} {Phys. Rev. B}\ }\textbf {\bibinfo {volume} {59}},\
  \bibinfo {pages} {1758} (\bibinfo {year} {1999})}\BibitemShut {NoStop}%
\bibitem [{\citenamefont {Blöchl}(1994)}]{blochl1994projector}%
  \BibitemOpen
  \bibfield  {author} {\bibinfo {author} {\bibfnamefont {P.~E.}\ \bibnamefont
  {Blöchl}},\ }\href {\doibase 10.1103/PhysRevB.50.17953} {\bibfield
  {journal} {\bibinfo  {journal} {Phys. Rev. B}\ }\textbf {\bibinfo {volume}
  {50}},\ \bibinfo {pages} {17953} (\bibinfo {year} {1994})}\BibitemShut
  {NoStop}%
\bibitem [{\citenamefont {Giannozzi}\ \emph
  {et~al.}(2009{\natexlab{b}})\citenamefont {Giannozzi}, \citenamefont
  {Baroni}, \citenamefont {Bonini}, \citenamefont {Calandra}, \citenamefont
  {Car}, \citenamefont {Cavazzoni}, \citenamefont {Ceresoli}, \citenamefont
  {Chiarotti}, \citenamefont {Cococcioni}, \citenamefont {Dabo}, \citenamefont
  {Corso}, \citenamefont {Gironcoli}, \citenamefont {Fabris}, \citenamefont
  {Fratesi}, \citenamefont {Gebauer}, \citenamefont {Gerstmann}, \citenamefont
  {Gougoussis}, \citenamefont {Kokalj}, \citenamefont {Lazzeri}, \citenamefont
  {Martin-Samos}, \citenamefont {Marzari}, \citenamefont {Mauri}, \citenamefont
  {Mazzarello}, \citenamefont {Paolini}, \citenamefont {Pasquarello},
  \citenamefont {Paulatto}, \citenamefont {Sbraccia}, \citenamefont {Scandolo},
  \citenamefont {Sclauzero}, \citenamefont {Seitsonen}, \citenamefont
  {Smogunov}, \citenamefont {Umari},\ and\ \citenamefont
  {Wentzcovitch}}]{giannozzi2009quantum}%
  \BibitemOpen
  \bibfield  {author} {\bibinfo {author} {\bibfnamefont {P.}~\bibnamefont
  {Giannozzi}}, \bibinfo {author} {\bibfnamefont {S.}~\bibnamefont {Baroni}},
  \bibinfo {author} {\bibfnamefont {N.}~\bibnamefont {Bonini}}, \bibinfo
  {author} {\bibfnamefont {M.}~\bibnamefont {Calandra}}, \bibinfo {author}
  {\bibfnamefont {R.}~\bibnamefont {Car}}, \bibinfo {author} {\bibfnamefont
  {C.}~\bibnamefont {Cavazzoni}}, \bibinfo {author} {\bibfnamefont
  {D.}~\bibnamefont {Ceresoli}}, \bibinfo {author} {\bibfnamefont {G.~L.}\
  \bibnamefont {Chiarotti}}, \bibinfo {author} {\bibfnamefont {M.}~\bibnamefont
  {Cococcioni}}, \bibinfo {author} {\bibfnamefont {I.}~\bibnamefont {Dabo}},
  \bibinfo {author} {\bibfnamefont {A.~D.}\ \bibnamefont {Corso}}, \bibinfo
  {author} {\bibfnamefont {S.~d.}\ \bibnamefont {Gironcoli}}, \bibinfo {author}
  {\bibfnamefont {S.}~\bibnamefont {Fabris}}, \bibinfo {author} {\bibfnamefont
  {G.}~\bibnamefont {Fratesi}}, \bibinfo {author} {\bibfnamefont
  {R.}~\bibnamefont {Gebauer}}, \bibinfo {author} {\bibfnamefont
  {U.}~\bibnamefont {Gerstmann}}, \bibinfo {author} {\bibfnamefont
  {C.}~\bibnamefont {Gougoussis}}, \bibinfo {author} {\bibfnamefont
  {A.}~\bibnamefont {Kokalj}}, \bibinfo {author} {\bibfnamefont
  {M.}~\bibnamefont {Lazzeri}}, \bibinfo {author} {\bibfnamefont
  {L.}~\bibnamefont {Martin-Samos}}, \bibinfo {author} {\bibfnamefont
  {N.}~\bibnamefont {Marzari}}, \bibinfo {author} {\bibfnamefont
  {F.}~\bibnamefont {Mauri}}, \bibinfo {author} {\bibfnamefont
  {R.}~\bibnamefont {Mazzarello}}, \bibinfo {author} {\bibfnamefont
  {S.}~\bibnamefont {Paolini}}, \bibinfo {author} {\bibfnamefont
  {A.}~\bibnamefont {Pasquarello}}, \bibinfo {author} {\bibfnamefont
  {L.}~\bibnamefont {Paulatto}}, \bibinfo {author} {\bibfnamefont
  {C.}~\bibnamefont {Sbraccia}}, \bibinfo {author} {\bibfnamefont
  {S.}~\bibnamefont {Scandolo}}, \bibinfo {author} {\bibfnamefont
  {G.}~\bibnamefont {Sclauzero}}, \bibinfo {author} {\bibfnamefont {A.~P.}\
  \bibnamefont {Seitsonen}}, \bibinfo {author} {\bibfnamefont {A.}~\bibnamefont
  {Smogunov}}, \bibinfo {author} {\bibfnamefont {P.}~\bibnamefont {Umari}}, \
  and\ \bibinfo {author} {\bibfnamefont {R.~M.}\ \bibnamefont {Wentzcovitch}},\
  }\href {\doibase 10.1088/0953-8984/21/39/395502} {\bibfield  {journal}
  {\bibinfo  {journal} {J. Phys.: Condens. Matter}\ }\textbf {\bibinfo {volume}
  {21}},\ \bibinfo {pages} {395502} (\bibinfo {year}
  {2009}{\natexlab{b}})}\BibitemShut {NoStop}%
\bibitem [{\citenamefont {Bartók}\ \emph {et~al.}(2010)\citenamefont
  {Bartók}, \citenamefont {Payne}, \citenamefont {Kondor},\ and\ \citenamefont
  {Csányi}}]{bartok2010gaussian}%
  \BibitemOpen
  \bibfield  {author} {\bibinfo {author} {\bibfnamefont {A.~P.}\ \bibnamefont
  {Bartók}}, \bibinfo {author} {\bibfnamefont {M.~C.}\ \bibnamefont {Payne}},
  \bibinfo {author} {\bibfnamefont {R.}~\bibnamefont {Kondor}}, \ and\ \bibinfo
  {author} {\bibfnamefont {G.}~\bibnamefont {Csányi}},\ }\href {\doibase
  10.1103/PhysRevLett.104.136403} {\bibfield  {journal} {\bibinfo  {journal}
  {Phys. Rev. Lett.}\ }\textbf {\bibinfo {volume} {104}},\ \bibinfo {pages}
  {136403} (\bibinfo {year} {2010})}\BibitemShut {NoStop}%
\bibitem [{\citenamefont {Powell}(1964)}]{P64}%
  \BibitemOpen
  \bibfield  {author} {\bibinfo {author} {\bibfnamefont {M.~J.~D.}\
  \bibnamefont {Powell}},\ }\href {\doibase 10.1093/comjnl/7.2.155} {\bibfield
  {journal} {\bibinfo  {journal} {The Computer Journal}\ }\textbf {\bibinfo
  {volume} {7}},\ \bibinfo {pages} {155} (\bibinfo {year} {1964})}\BibitemShut
  {NoStop}%
\bibitem [{\citenamefont {Wang}\ \emph {et~al.}(2004)\citenamefont {Wang},
  \citenamefont {Wolf}, \citenamefont {Caldwell}, \citenamefont {Kollman},\
  and\ \citenamefont {Case}}]{GAFF}%
  \BibitemOpen
  \bibfield  {author} {\bibinfo {author} {\bibfnamefont {J.}~\bibnamefont
  {Wang}}, \bibinfo {author} {\bibfnamefont {R.~M.}\ \bibnamefont {Wolf}},
  \bibinfo {author} {\bibfnamefont {J.~W.}\ \bibnamefont {Caldwell}}, \bibinfo
  {author} {\bibfnamefont {P.~A.}\ \bibnamefont {Kollman}}, \ and\ \bibinfo
  {author} {\bibfnamefont {D.~A.}\ \bibnamefont {Case}},\ }\href {\doibase
  10.1002/jcc.20035} {\bibfield  {journal} {\bibinfo  {journal} {Journal of
  Computational Chemistry}\ }\textbf {\bibinfo {volume} {25}},\ \bibinfo
  {pages} {1157} (\bibinfo {year} {2004})}\BibitemShut {NoStop}%
\bibitem [{\citenamefont {Tuckerman}\ \emph {et~al.}(2000)\citenamefont
  {Tuckerman}, \citenamefont {Yarne}, \citenamefont {Samuelson}, \citenamefont
  {Hughes},\ and\ \citenamefont {Martyna}}]{PINYMD}%
  \BibitemOpen
  \bibfield  {author} {\bibinfo {author} {\bibfnamefont {M.~E.}\ \bibnamefont
  {Tuckerman}}, \bibinfo {author} {\bibfnamefont {D.}~\bibnamefont {Yarne}},
  \bibinfo {author} {\bibfnamefont {S.~O.}\ \bibnamefont {Samuelson}}, \bibinfo
  {author} {\bibfnamefont {A.~L.}\ \bibnamefont {Hughes}}, \ and\ \bibinfo
  {author} {\bibfnamefont {G.~J.}\ \bibnamefont {Martyna}},\ }\href {\doibase
  http://dx.doi.org/10.1016/S0010-4655(00)00077-1} {\bibfield  {journal}
  {\bibinfo  {journal} {Computer Physics Communications}\ }\textbf {\bibinfo
  {volume} {128}},\ \bibinfo {pages} {333 } (\bibinfo {year}
  {2000})}\BibitemShut {NoStop}%
\bibitem [{\citenamefont {Perdew}\ and\ \citenamefont {Zunger}(1981)}]{PZLDA}%
  \BibitemOpen
  \bibfield  {author} {\bibinfo {author} {\bibfnamefont {J.~P.}\ \bibnamefont
  {Perdew}}\ and\ \bibinfo {author} {\bibfnamefont {A.}~\bibnamefont
  {Zunger}},\ }\href {\doibase 10.1103/PhysRevB.23.5048} {\bibfield  {journal}
  {\bibinfo  {journal} {Phys. Rev. B}\ }\textbf {\bibinfo {volume} {23}},\
  \bibinfo {pages} {5048} (\bibinfo {year} {1981})}\BibitemShut {NoStop}%
\bibitem [{\citenamefont {Tuckerman}\ and\ \citenamefont
  {Marx}(2001)}]{MT_PRL}%
  \BibitemOpen
  \bibfield  {author} {\bibinfo {author} {\bibfnamefont {M.~E.}\ \bibnamefont
  {Tuckerman}}\ and\ \bibinfo {author} {\bibfnamefont {D.}~\bibnamefont
  {Marx}},\ }\href {\doibase 10.1103/PhysRevLett.86.4946} {\bibfield  {journal}
  {\bibinfo  {journal} {Phys. Rev. Lett.}\ }\textbf {\bibinfo {volume} {86}},\
  \bibinfo {pages} {4946} (\bibinfo {year} {2001})}\BibitemShut {NoStop}%
\bibitem [{\citenamefont {Vapnik}(2000)}]{vapnik2000nature}%
  \BibitemOpen
  \bibfield  {author} {\bibinfo {author} {\bibfnamefont {V.}~\bibnamefont
  {Vapnik}},\ }\href {http://books.google.de/books?id=sna9BaxVbj8C} {\emph
  {\bibinfo {title} {The Nature of Statistical Learning Theory}}},\ Information
  Science and Statistics\ (\bibinfo  {publisher} {Springer},\ \bibinfo {year}
  {2000})\BibitemShut {NoStop}%
\bibitem [{\citenamefont {Cangi}\ \emph {et~al.}(2011)\citenamefont {Cangi},
  \citenamefont {Lee}, \citenamefont {Elliott}, \citenamefont {Burke},\ and\
  \citenamefont {Gross}}]{CLEB11}%
  \BibitemOpen
  \bibfield  {author} {\bibinfo {author} {\bibfnamefont {A.}~\bibnamefont
  {Cangi}}, \bibinfo {author} {\bibfnamefont {D.}~\bibnamefont {Lee}}, \bibinfo
  {author} {\bibfnamefont {P.}~\bibnamefont {Elliott}}, \bibinfo {author}
  {\bibfnamefont {K.}~\bibnamefont {Burke}}, \ and\ \bibinfo {author}
  {\bibfnamefont {E.~K.~U.}\ \bibnamefont {Gross}},\ }\href {\doibase
  10.1103/PhysRevLett.106.236404} {\bibfield  {journal} {\bibinfo  {journal}
  {Phys. Rev. Lett.}\ }\textbf {\bibinfo {volume} {106}},\ \bibinfo {pages}
  {236404} (\bibinfo {year} {2011})}\BibitemShut {NoStop}%
\bibitem [{\citenamefont {Hohenberg}\ and\ \citenamefont {Kohn}(1964)}]{HK64}%
  \BibitemOpen
  \bibfield  {author} {\bibinfo {author} {\bibfnamefont {P.}~\bibnamefont
  {Hohenberg}}\ and\ \bibinfo {author} {\bibfnamefont {W.}~\bibnamefont
  {Kohn}},\ }\href {\doibase 10.1103/PhysRev.136.B864} {\bibfield  {journal}
  {\bibinfo  {journal} {Phys. Rev.}\ }\textbf {\bibinfo {volume} {136}},\
  \bibinfo {pages} {B864} (\bibinfo {year} {1964})}\BibitemShut {NoStop}%
\bibitem [{\citenamefont {Troullier}\ and\ \citenamefont
  {Martins}(1991)}]{TM91}%
  \BibitemOpen
  \bibfield  {author} {\bibinfo {author} {\bibfnamefont {N.}~\bibnamefont
  {Troullier}}\ and\ \bibinfo {author} {\bibfnamefont {J.~L.}\ \bibnamefont
  {Martins}},\ }\href {\doibase 10.1103/PhysRevB.43.1993} {\bibfield  {journal}
  {\bibinfo  {journal} {Phys. Rev. B}\ }\textbf {\bibinfo {volume} {43}},\
  \bibinfo {pages} {1993} (\bibinfo {year} {1991})}\BibitemShut {NoStop}%
\bibitem [{\citenamefont {Hastie}\ \emph {et~al.}(2009)\citenamefont {Hastie},
  \citenamefont {Tibshirani},\ and\ \citenamefont
  {Friedman}}]{hastie2009elements}%
  \BibitemOpen
  \bibfield  {author} {\bibinfo {author} {\bibfnamefont {T.}~\bibnamefont
  {Hastie}}, \bibinfo {author} {\bibfnamefont {R.}~\bibnamefont {Tibshirani}},
  \ and\ \bibinfo {author} {\bibfnamefont {J.}~\bibnamefont {Friedman}},\
  }\href {http://www.springer.com/computer/ai/book/978-0-387-84857-0} {\emph
  {\bibinfo {title} {The Elements of Statistical Learning - Data Mining,
  Inference, and Prediction}}},\ \bibinfo {edition} {2nd}\ ed.,\ Springer
  Series in Statistics\ (\bibinfo  {publisher} {Springer},\ \bibinfo {year}
  {2009})\BibitemShut {NoStop}%
\bibitem [{\citenamefont {Vu}\ \emph {et~al.}(2015)\citenamefont {Vu},
  \citenamefont {Snyder}, \citenamefont {Li}, \citenamefont {Rupp},
  \citenamefont {Chen}, \citenamefont {Khelif}, \citenamefont {Müller},\ and\
  \citenamefont {Burke}}]{vu2015understanding}%
  \BibitemOpen
  \bibfield  {author} {\bibinfo {author} {\bibfnamefont {K.}~\bibnamefont
  {Vu}}, \bibinfo {author} {\bibfnamefont {J.~C.}\ \bibnamefont {Snyder}},
  \bibinfo {author} {\bibfnamefont {L.}~\bibnamefont {Li}}, \bibinfo {author}
  {\bibfnamefont {M.}~\bibnamefont {Rupp}}, \bibinfo {author} {\bibfnamefont
  {B.~F.}\ \bibnamefont {Chen}}, \bibinfo {author} {\bibfnamefont
  {T.}~\bibnamefont {Khelif}}, \bibinfo {author} {\bibfnamefont {K.-R.}\
  \bibnamefont {Müller}}, \ and\ \bibinfo {author} {\bibfnamefont
  {K.}~\bibnamefont {Burke}},\ }\href {\doibase 10.1002/qua.24939} {\bibfield
  {journal} {\bibinfo  {journal} {International Journal of Quantum Chemistry}\
  }\textbf {\bibinfo {volume} {115}},\ \bibinfo {pages} {1115} (\bibinfo {year}
  {2015})}\BibitemShut {NoStop}%
\bibitem [{\citenamefont {Werner}\ \emph {et~al.}(2015)\citenamefont {Werner},
  \citenamefont {Knowles}, \citenamefont {Knizia}, \citenamefont {Manby},
  \citenamefont {{Schütz}} \emph {et~al.}}]{MOLPRO_brief}%
  \BibitemOpen
  \bibfield  {author} {\bibinfo {author} {\bibfnamefont {H.-J.}\ \bibnamefont
  {Werner}}, \bibinfo {author} {\bibfnamefont {P.~J.}\ \bibnamefont {Knowles}},
  \bibinfo {author} {\bibfnamefont {G.}~\bibnamefont {Knizia}}, \bibinfo
  {author} {\bibfnamefont {F.~R.}\ \bibnamefont {Manby}}, \bibinfo {author}
  {\bibfnamefont {M.}~\bibnamefont {{Schütz}}},  \emph {et~al.},\ }\href@noop
  {} {\enquote {\bibinfo {title} {Molpro, version 2015.1, a package of ab
  initio programs},}\ } (\bibinfo {year} {2015})\BibitemShut {NoStop}%
\bibitem [{\citenamefont {Adler}\ \emph {et~al.}(2007)\citenamefont {Adler},
  \citenamefont {Knizia},\ and\ \citenamefont {Werner}}]{AKW07}%
  \BibitemOpen
  \bibfield  {author} {\bibinfo {author} {\bibfnamefont {T.~B.}\ \bibnamefont
  {Adler}}, \bibinfo {author} {\bibfnamefont {G.}~\bibnamefont {Knizia}}, \
  and\ \bibinfo {author} {\bibfnamefont {H.-J.}\ \bibnamefont {Werner}},\
  }\href@noop {} {\bibfield  {journal} {\bibinfo  {journal} {Journal of
  Chemical Physics}\ }\textbf {\bibinfo {volume} {127}},\ \bibinfo {pages}
  {221106} (\bibinfo {year} {2007})}\BibitemShut {NoStop}%
\bibitem [{\citenamefont {Wang}\ \emph {et~al.}(2001)\citenamefont {Wang},
  \citenamefont {Wang}, \citenamefont {Kollman},\ and\ \citenamefont
  {Case}}]{antechamber}%
  \BibitemOpen
  \bibfield  {author} {\bibinfo {author} {\bibfnamefont {J.}~\bibnamefont
  {Wang}}, \bibinfo {author} {\bibfnamefont {W.}~\bibnamefont {Wang}}, \bibinfo
  {author} {\bibfnamefont {P.~A.}\ \bibnamefont {Kollman}}, \ and\ \bibinfo
  {author} {\bibfnamefont {D.~A.}\ \bibnamefont {Case}},\ }\href@noop {}
  {\bibfield  {journal} {\bibinfo  {journal} {Journal of the American Chemical
  Society}\ }\textbf {\bibinfo {volume} {222}},\ \bibinfo {pages} {U403}
  (\bibinfo {year} {2001})}\BibitemShut {NoStop}%
\bibitem [{\citenamefont {Frisch}\ \emph {et~al.}(2009)\citenamefont {Frisch},
  \citenamefont {Trucks}, \citenamefont {Schlegel}, \citenamefont {Scuseria},
  \citenamefont {Robb}, \citenamefont {Cheeseman}, \citenamefont {Scalmani},
  \citenamefont {Barone}, \citenamefont {Mennucci}, \citenamefont {Petersson}
  \emph {et~al.}}]{Gaussian09}%
  \BibitemOpen
  \bibfield  {author} {\bibinfo {author} {\bibfnamefont {M.}~\bibnamefont
  {Frisch}}, \bibinfo {author} {\bibfnamefont {G.}~\bibnamefont {Trucks}},
  \bibinfo {author} {\bibfnamefont {H.~B.}\ \bibnamefont {Schlegel}}, \bibinfo
  {author} {\bibfnamefont {G.}~\bibnamefont {Scuseria}}, \bibinfo {author}
  {\bibfnamefont {M.}~\bibnamefont {Robb}}, \bibinfo {author} {\bibfnamefont
  {J.}~\bibnamefont {Cheeseman}}, \bibinfo {author} {\bibfnamefont
  {G.}~\bibnamefont {Scalmani}}, \bibinfo {author} {\bibfnamefont
  {V.}~\bibnamefont {Barone}}, \bibinfo {author} {\bibfnamefont
  {B.}~\bibnamefont {Mennucci}}, \bibinfo {author} {\bibfnamefont
  {G.}~\bibnamefont {Petersson}},  \emph {et~al.},\ }\href@noop {} {\enquote
  {\bibinfo {title} {Gaussian 09},}\ } (\bibinfo {year} {2009})\BibitemShut
  {NoStop}%
\bibitem [{\citenamefont {Bayly}\ \emph {et~al.}(1993)\citenamefont {Bayly},
  \citenamefont {Cieplak}, \citenamefont {Cornell},\ and\ \citenamefont
  {Kollman}}]{RESP}%
  \BibitemOpen
  \bibfield  {author} {\bibinfo {author} {\bibfnamefont {C.~I.}\ \bibnamefont
  {Bayly}}, \bibinfo {author} {\bibfnamefont {P.}~\bibnamefont {Cieplak}},
  \bibinfo {author} {\bibfnamefont {W.}~\bibnamefont {Cornell}}, \ and\
  \bibinfo {author} {\bibfnamefont {P.~A.}\ \bibnamefont {Kollman}},\ }\href
  {\doibase 10.1021/j100142a004} {\bibfield  {journal} {\bibinfo  {journal}
  {The Journal of Physical Chemistry}\ }\textbf {\bibinfo {volume} {97}},\
  \bibinfo {pages} {10269} (\bibinfo {year} {1993})}\BibitemShut {NoStop}%
\bibitem [{\citenamefont {Martyna}\ \emph {et~al.}(1992)\citenamefont
  {Martyna}, \citenamefont {Klein},\ and\ \citenamefont {Tuckerman}}]{nhc}%
  \BibitemOpen
  \bibfield  {author} {\bibinfo {author} {\bibfnamefont {G.~J.}\ \bibnamefont
  {Martyna}}, \bibinfo {author} {\bibfnamefont {M.~L.}\ \bibnamefont {Klein}},
  \ and\ \bibinfo {author} {\bibfnamefont {M.}~\bibnamefont {Tuckerman}},\
  }\href {\doibase http://dx.doi.org/10.1063/1.463940} {\bibfield  {journal}
  {\bibinfo  {journal} {The Journal of Chemical Physics}\ }\textbf {\bibinfo
  {volume} {97}},\ \bibinfo {pages} {2635} (\bibinfo {year}
  {1992})}\BibitemShut {NoStop}%
\bibitem [{\citenamefont {Tuckerman}\ \emph {et~al.}(1992)\citenamefont
  {Tuckerman}, \citenamefont {Berne},\ and\ \citenamefont {Martyna}}]{respa}%
  \BibitemOpen
  \bibfield  {author} {\bibinfo {author} {\bibfnamefont {M.}~\bibnamefont
  {Tuckerman}}, \bibinfo {author} {\bibfnamefont {B.~J.}\ \bibnamefont
  {Berne}}, \ and\ \bibinfo {author} {\bibfnamefont {G.~J.}\ \bibnamefont
  {Martyna}},\ }\href {\doibase http://dx.doi.org/10.1063/1.463137} {\bibfield
  {journal} {\bibinfo  {journal} {The Journal of Chemical Physics}\ }\textbf
  {\bibinfo {volume} {97}},\ \bibinfo {pages} {1990} (\bibinfo {year}
  {1992})}\BibitemShut {NoStop}%
\bibitem [{\citenamefont {VandeVondele}\ \emph {et~al.}(2005)\citenamefont
  {VandeVondele}, \citenamefont {Krack}, \citenamefont {Mohamed}, \citenamefont
  {Parrinello}, \citenamefont {Chassaing},\ and\ \citenamefont {Hutter}}]{QS}%
  \BibitemOpen
  \bibfield  {author} {\bibinfo {author} {\bibfnamefont {J.}~\bibnamefont
  {VandeVondele}}, \bibinfo {author} {\bibfnamefont {M.}~\bibnamefont {Krack}},
  \bibinfo {author} {\bibfnamefont {F.}~\bibnamefont {Mohamed}}, \bibinfo
  {author} {\bibfnamefont {M.}~\bibnamefont {Parrinello}}, \bibinfo {author}
  {\bibfnamefont {T.}~\bibnamefont {Chassaing}}, \ and\ \bibinfo {author}
  {\bibfnamefont {J.}~\bibnamefont {Hutter}},\ }\href@noop {} {\bibfield
  {journal} {\bibinfo  {journal} {Computer Physics Communications}\ }\textbf
  {\bibinfo {volume} {167}},\ \bibinfo {pages} {103} (\bibinfo {year}
  {2005})}\BibitemShut {NoStop}%
\bibitem [{\citenamefont {Hutter}\ \emph {et~al.}(2014)\citenamefont {Hutter},
  \citenamefont {Iannuzzi}, \citenamefont {Schiffmann},\ and\ \citenamefont
  {VandeVondele}}]{CP2K}%
  \BibitemOpen
  \bibfield  {author} {\bibinfo {author} {\bibfnamefont {J.}~\bibnamefont
  {Hutter}}, \bibinfo {author} {\bibfnamefont {M.}~\bibnamefont {Iannuzzi}},
  \bibinfo {author} {\bibfnamefont {F.}~\bibnamefont {Schiffmann}}, \ and\
  \bibinfo {author} {\bibfnamefont {J.}~\bibnamefont {VandeVondele}},\
  }\href@noop {} {\bibfield  {journal} {\bibinfo  {journal} {Wiley
  Interdisciplinary Reviews: Computational Molecular Science}\ }\textbf
  {\bibinfo {volume} {4}},\ \bibinfo {pages} {15} (\bibinfo {year}
  {2014})}\BibitemShut {NoStop}%
\bibitem [{\citenamefont {Lippert}\ \emph {et~al.}(2010)\citenamefont
  {Lippert}, \citenamefont {Hutter},\ and\ \citenamefont {Parrinello}}]{GPW}%
  \BibitemOpen
  \bibfield  {author} {\bibinfo {author} {\bibfnamefont {G.}~\bibnamefont
  {Lippert}}, \bibinfo {author} {\bibfnamefont {J.}~\bibnamefont {Hutter}}, \
  and\ \bibinfo {author} {\bibfnamefont {M.}~\bibnamefont {Parrinello}},\
  }\href@noop {} {\bibfield  {journal} {\bibinfo  {journal} {Molecular
  Physics}\ }\textbf {\bibinfo {volume} {92}},\ \bibinfo {pages} {477}
  (\bibinfo {year} {2010})}\BibitemShut {NoStop}%
\bibitem [{\citenamefont {VandeVondele}\ and\ \citenamefont
  {Hutter}(2007)}]{MOLOPT}%
  \BibitemOpen
  \bibfield  {author} {\bibinfo {author} {\bibfnamefont {J.}~\bibnamefont
  {VandeVondele}}\ and\ \bibinfo {author} {\bibfnamefont {J.}~\bibnamefont
  {Hutter}},\ }\href@noop {} {\bibfield  {journal} {\bibinfo  {journal} {The
  Journal of Chemical Physics}\ }\textbf {\bibinfo {volume} {127}},\ \bibinfo
  {pages} {114105} (\bibinfo {year} {2007})}\BibitemShut {NoStop}%
\bibitem [{\citenamefont {Goedecker}\ \emph {et~al.}(1996)\citenamefont
  {Goedecker}, \citenamefont {Teter},\ and\ \citenamefont {Hutter}}]{GTHpp}%
  \BibitemOpen
  \bibfield  {author} {\bibinfo {author} {\bibfnamefont {S.}~\bibnamefont
  {Goedecker}}, \bibinfo {author} {\bibfnamefont {M.}~\bibnamefont {Teter}}, \
  and\ \bibinfo {author} {\bibfnamefont {J.}~\bibnamefont {Hutter}},\ }\href
  {\doibase 10.1103/PhysRevB.54.1703} {\bibfield  {journal} {\bibinfo
  {journal} {Phys. Rev. B}\ }\textbf {\bibinfo {volume} {54}},\ \bibinfo
  {pages} {1703} (\bibinfo {year} {1996})}\BibitemShut {NoStop}%
\bibitem [{\citenamefont {Krack}(2005)}]{PBEpp}%
  \BibitemOpen
  \bibfield  {author} {\bibinfo {author} {\bibfnamefont {M.}~\bibnamefont
  {Krack}},\ }\href@noop {} {\bibfield  {journal} {\bibinfo  {journal}
  {Theoretica Chimica Acta}\ }\textbf {\bibinfo {volume} {114}},\ \bibinfo
  {pages} {145} (\bibinfo {year} {2005})}\BibitemShut {NoStop}%
\bibitem [{\citenamefont {VandeVondele}\ and\ \citenamefont
  {Hutter}(2003)}]{OT}%
  \BibitemOpen
  \bibfield  {author} {\bibinfo {author} {\bibfnamefont {J.}~\bibnamefont
  {VandeVondele}}\ and\ \bibinfo {author} {\bibfnamefont {J.}~\bibnamefont
  {Hutter}},\ }\href@noop {} {\bibfield  {journal} {\bibinfo  {journal} {The
  Journal of Chemical Physics}\ }\textbf {\bibinfo {volume} {118}},\ \bibinfo
  {pages} {4365} (\bibinfo {year} {2003})}\BibitemShut {NoStop}%
\bibitem [{\citenamefont {Bahn}\ and\ \citenamefont {Jacobsen}(2002)}]{BJ02}%
  \BibitemOpen
  \bibfield  {author} {\bibinfo {author} {\bibfnamefont {S.~R.}\ \bibnamefont
  {Bahn}}\ and\ \bibinfo {author} {\bibfnamefont {K.~W.}\ \bibnamefont
  {Jacobsen}},\ }\href {\doibase 10.1109/5992.998641} {\bibfield  {journal}
  {\bibinfo  {journal} {Comput. Sci. Eng.}\ }\textbf {\bibinfo {volume} {4}},\
  \bibinfo {pages} {56} (\bibinfo {year} {2002})}\BibitemShut {NoStop}%
\end{thebibliography}%
\end{document}


\title{By-passing the Kohn-Sham equations with machine learning\\Supplemental Information}

\author{Felix Brockherde}
\affiliation{Machine Learning Group, Technische Universität Berlin, Marchstr. 23, 10587 Berlin, Germany}
\affiliation{Max-Planck-Institut für Mikrostrukturphysik, Weinberg 2, 06120 Halle, Germany}
\author{Leslie Vogt}
\affiliation{Department of Chemistry, New York University, New York, NY 10003, USA}
\author{Li Li}
\affiliation{Departments of Physics and Astronomy, University of California, Irvine, CA 92697, USA}
\author{Mark E. Tuckerman}
\affiliation{Department of Chemistry, New York University, New York, NY 10003, USA}
\affiliation{Courant Institute of Mathematical Science, New York University, New York, NY 10003, USA}
\affiliation{NYU-ECNU Center for Computational Chemistry at NYU Shanghai, 3663 Zhongshan Road North, Shanghai 200062, China}
\author{Kieron Burke}
\thanks{to whom correspondence should be addressed.}
\affiliation{Departments of Chemistry, University of California, Irvine, CA 92697, USA}
\affiliation{Departments of Physics and Astronomy, University of California, Irvine, CA 92697, USA}
\author{Klaus-Robert Müller}
\thanks{to whom correspondence should be addressed.}
\affiliation{Machine Learning Group, Technische Universität Berlin, Marchstr. 23, 10587 Berlin, Germany}
\affiliation{Department of Brain and Cognitive Engineering, Korea University, Anam-dong, Seongbuk-gu, Seoul 136--713, Republic of Korea}
\affiliation{Max Planck Institute for Informatics, Stuhlsatzenhausweg, 66123 Saarbrücken, Germany}

\date{\today}

\maketitle

\section{Kernel Ridge Regression}
Kernel Ridge Regression\citep{hastie2009elements,vu2015understanding} (KRR) is a machine learning method for regression. We introduce the method for abstract training points $(x_i, y_i)$, i.e.~features $x_1, \dots x_M \in \mathbb{R}^d$ and associated labels $\mathbf{Y} = {(y_1, \dots, y_M)}^T \in \mathbb{R}^M$ and describe the actual models used in the main text afterwards. We want to model a function $f: \mathbb{R}^d \rightarrow \mathbb{R}$ that maps from features to labels. This model should not be `learned by heart' but perform well on unseen data (i.e.~\textit{generalize}). We first restrict the set of possible functions to the reproducing kernel Hilbert space (RKHS) $\mathcal{H}$ on the space of discretized densities that is induced by the Gaussian kernel function

\begin{align}
    k(x, x') = \exp\left(-\frac{||x - x'||^2}{2\sigma^2}\right).
\end{align}
The restriction is very mild and rather technical; more interesting is the choice of the kernel function which determines the scalar product (and thus the norm) of the RKHS\@. Leaving rigor aside, the Gaussian kernel induces an RKHS norm $||f||_{\mathcal{H}}$ that is smaller for simpler, smoother functions and higher for more complicated, oscillating functions. We minimize the empirical risk functional

\begin{align} \label{empirical-risk-functional}
\mathcal{C}(f) = \sum_{i=1}^M |y_i - f(x_i)|^2 + \lambda \lVert f\rVert^2_\mathcal{H}
\end{align}
that defines a trade-off between error on the training points and smoothness of the function controlled by the hyper-parameter $\lambda$.

The representer theorem\citep{Schoelkopf2001} allows us to assume that the solution to Eq.~\ref{empirical-risk-functional} is given by a linear combination of kernel functions $f = \sum_{i=1}^M\boldsymbol{\alpha}_i k(x_i, \cdot)$. It now suffices to solve

\begin{align}
    \mathcal{C}(\boldsymbol{\alpha}) &= \sum_{i=1}^M |y_i - f(x_i)|^2 + \lambda \lVert f\rVert^2_\mathcal{H}\\
    &= \sum_{i=1}^M |y_i - f(x_i)|^2 + \lambda \boldsymbol{\alpha}^\intercal \mathbf{K} \boldsymbol{\alpha},
\end{align}
where $\mathbf{K}_{ij} = k(x_i, x_j)$ is the kernel matrix. The solution is given by

\begin{align} \label{eq:krr-analytical}
    \boldsymbol{\alpha} = {\left(\mathbf{K} + \lambda \mathbf{I}\right)}^{-1} \mathbf{Y}.
\end{align}

Note that all model parameters and hyper-parameters are estimated on the training set; the  hyper-parameter choice makes use of standard cross-validation procedures (see \citet{hansen2013}). Once the model is fixed after training, it is applied unchanged out-of-sample.

We use this method for various maps:
\paragraph{Non-interacting kinetic energy functional ($T\s^{\text{ML}}[n]$, 1-D).} The training points are given by pairs of densities and associated kinetic energies. We discretize the densities and use them in vectorial form, i.e.~$n \in \mathbb{R}^G$. Thus, the \textit{functional} $\mathcal{L}^2\rightarrow \mathbb{R}$ is modeled as a \textit{function} $\mathbb{R^G}\rightarrow \mathbb{R}$
\paragraph{ML-OF map (1-D).} The training points are given by pairs of discretized 1-D box potentials and associated total energies.
\paragraph{ML-KS map (3-D).} The training points are given by pairs of discretized Gaussians potentials (as described in the main text) and total energies.
\paragraph{Total energy functional ($E^\text{ML}[n]$, 3-D).} The training points are given by pairs of densities in basis function representation (see below) and associated total energies. Just as for $T\s^{\text{ML}}$, this \textit{functional} is modeled as a \textit{function}.

\section{ML Hohenberg-Kohn map}
The basis representation for the densities is given by

\begin{align}\label{density_basis}
    n(x) = \sum_{l=1}^L u^{(l)} \phi_l(x),
\end{align}
where $\phi_l$ are the $L$ basis functions. We introduce some notation and continue to write the density in grid representation as $n$, and its basis coefficients as $u$. We can then write the HK map model as

\begin{align}
    n^\text{ML}[v](x) = \sum_{l=1}^L u^{(l)}[v] \phi_l(x),
\end{align}
where the $L$ basis function coefficients are regular KRR models,

\begin{align}
    u^{(l)}[v] = \sum_{i=1}^M \beta_i^{(l)} k(v, v_i),
\end{align}
of external potentials $v$ with a Gaussian kernel function. The contribution of the error to the cost function can be formulated as

\begin{align}
e(\boldsymbol{\beta}) &= \sum_{i=1}^M \lVert n_i - n^\text{ML}[v_i] \rVert_{\mathcal{L}_2}^2\\
          &= \sum_{i=1}^M \left\lVert n_i - \sum_{l=1}^L \sum_{j=1}^M \beta_j^{(l)} k(v_i, v_j) \phi_l \right\rVert_{\mathcal{L}_2},
\end{align}
with the $\mathcal{L}_2$ norm.
We write this cost function in terms of basis function coefficients.
This can be viewed as projecting the inside of the
norm on each basis function. Assuming orthogonality of the basis functions yields

\begin{align} \label{eq:direct-model}
e(\boldsymbol{\beta}) = \sum_{i=1}^M \sum_{l=1}^L \left |u_i^{(l)} - \sum_{j=1}^M \beta_j^{(l)} k(v_i, v_j) \right|^2.
\end{align}
where $u_i^{(l)} = \langle n_i , \phi_l \rangle$ is
the $l$-th basis function coefficient of the $i$-th training density, as defined in Eq.~\ref{density_basis} if orthogonality is satisfied.
After reordering the sums over $i$ and $l$, we view each $l$ independently and solve analogously to regular KRR

\begin{align}
\boldsymbol{\beta}^{(l)} = {\left(\mathbf{K}_{\sigma^{(l)}} + \lambda^{(l)} \mathbf{I}\right)}^{-1} \mathbf{u}^{(l)}, \quad l = 1, \dots, L
\end{align}
where, for each basis function $l$, $\lambda^{(l)}$ is a regularization parameter, $\mathbf{K}_{\sigma^{(l)}}$ is a Gaussian kernel with kernel width $\sigma^{(l)}$.
The $\lambda^{(l)}$ and $\sigma^{(l)}$ can be chosen
individually for each basis function via independent cross-validation
(see \citep{muller2001introduction,hansen2013}).

\section{Basis functions}
\paragraph{Fourier basis.}
We define the basis as

\begin{align}
  \phi_l(x) =
  \begin{cases}
    \cos\left\{2\pi x(l-1)/2\right\}, &l \text{ odd}\\
    \sin\left\{2\pi xl/2\right\}, &l \text{ even}
  \end{cases}
  \quad l=1, \dots , L.
\end{align}
We transform the density efficiently via the discrete Fourier transform

\begin{align}
  u_i^{(l)} =
    \sum_{m=1}^G n_i(x_m) \phi_l(x_m).
\end{align}
The back-projection is written as

\begin{align}
    n^\text{ML}[v](x) = \sum_{l=1}^L u^{(l)}[v]\phi_l(x).
\end{align}
\paragraph{KPCA basis.}
We define the basis as:

\begin{align}
 \phi^{\mathrm{KPCA}}_l = \sum_{j=1}^M p_j^{(l)} \Phi(n_j).
\end{align}
The parameters $p_j^{(l)}$ are found by eigen-decomposition of the Kernel matrix. The KCPA basis coefficients are given by

\begin{align}
  u_i^{(l)} &= \langle\Phi(n_i), \phi_l^{\mathrm{KPCA}}
 \rangle = \sum_{j=1}^M p_j^{(l)} k(n_j, n_i)
\end{align}
with kernel map $\Phi$.
The back-projection for KPCA is not trivial but several solutions exist.
We follow \citet{bakir2004learning} and learn the back-projection map.

\section{Gradient descent issues}
There are two ways to remedy problems of the gradient descent procedure: First, the gradient descent step can be ``de-noised'' by projecting the gradient onto the data manifold and thus removing the noisy directions. Secondly, the directions outside of the data manifold can be removed in a preprocessing step to get rid of the influence of the noisy directions on the gradient completely. Both methods yield similar results.

Several approaches exist for describing and projecting onto the data manifold. Common to each approach is the idea to find principle components and to project on those in which direction the densities have largest variance. Best results are reported \citep{snyder2015nonlinear} by using Kernel Principle Component Analysis\citep{SSM98} (KPCA), a non-linear generalization of PCA\@.

%
%
There are three issues with the assumed gradient-based approaches: First, the correct choice of the number of (K)PCA components $K$ has to be made. It is generally possible to view it as a hyper-parameter and find the optimal $K$ via cross-validation. However, we can not choose fractional $K$s. One $K$ might be not enough and $K+1$ too much information.
Second, the data points only lie in a bounded region of a manifold that can be described via PCA components.
It is still possible for the gradient descent to walk outside this bounded region toward a point where the model has no information and thus the gradients become inaccurate.
A (K)PCA method that only accesses the scalar products between points in the data set can not solve this\citep{kiraly2014learning}.
Third, it might not be possible to find a suitable pre-image for a ground-state density given by (K)PCA coefficients\citep{inputspacevs}.

\section{Molecular Datasets}
\begin{figure}[h]
\includegraphics{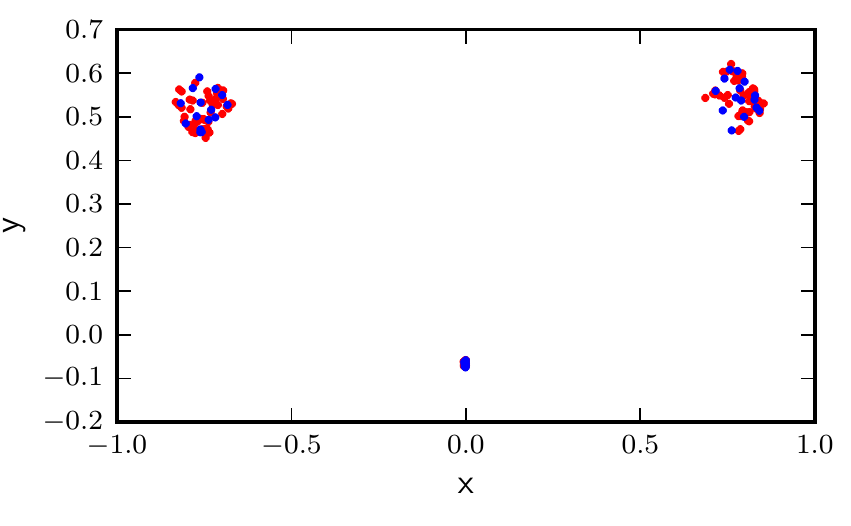}
\caption{\label{fig:h2o-dataset} The extent of the $\mathsf{H_2O}$ dataset. The figure shows the atom coordinates in angstrom. Blue are atoms from $15$ training points, red from $50$ test points.
}
\end{figure}

The extent of the dataset for $\mathsf{H_2O}$ is visualized in Fig.~\ref{fig:h2o-dataset}. In this case, conformers were generated from random displacements from the optimized geometry.

\begin{figure}[h]
\includegraphics{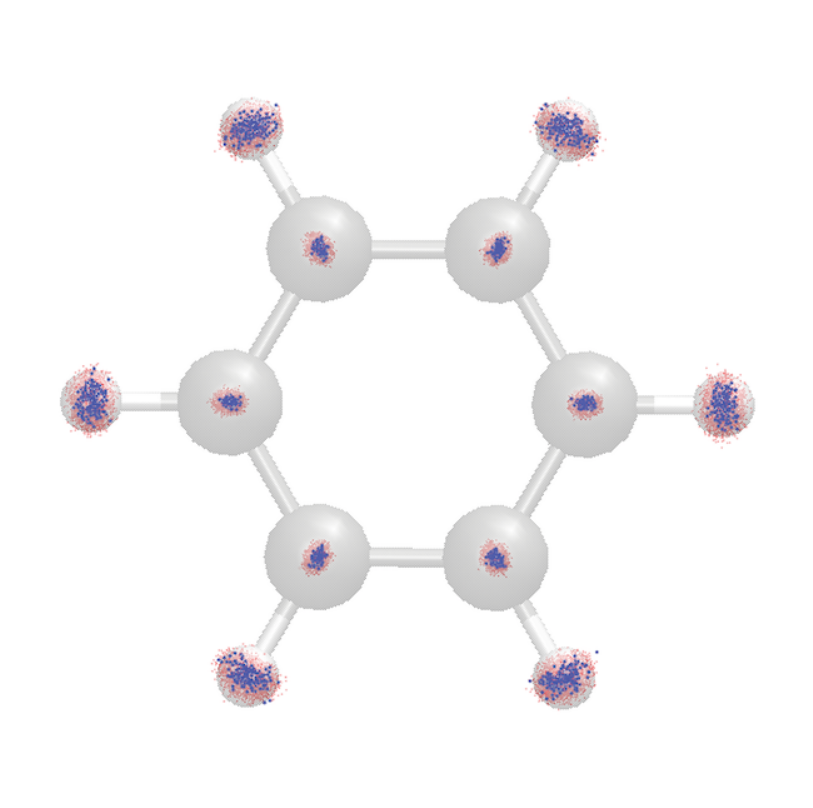}
\caption{\label{fig:benzene-dataset} The extent of the benzene conformers generated by MD (red points). K-means sampling is used to select 2,000 representative points. Test points from an independent trajectory are in blue and are offset from the molecular plane for clarity.}
\end{figure}

\begin{figure}[h]
\includegraphics{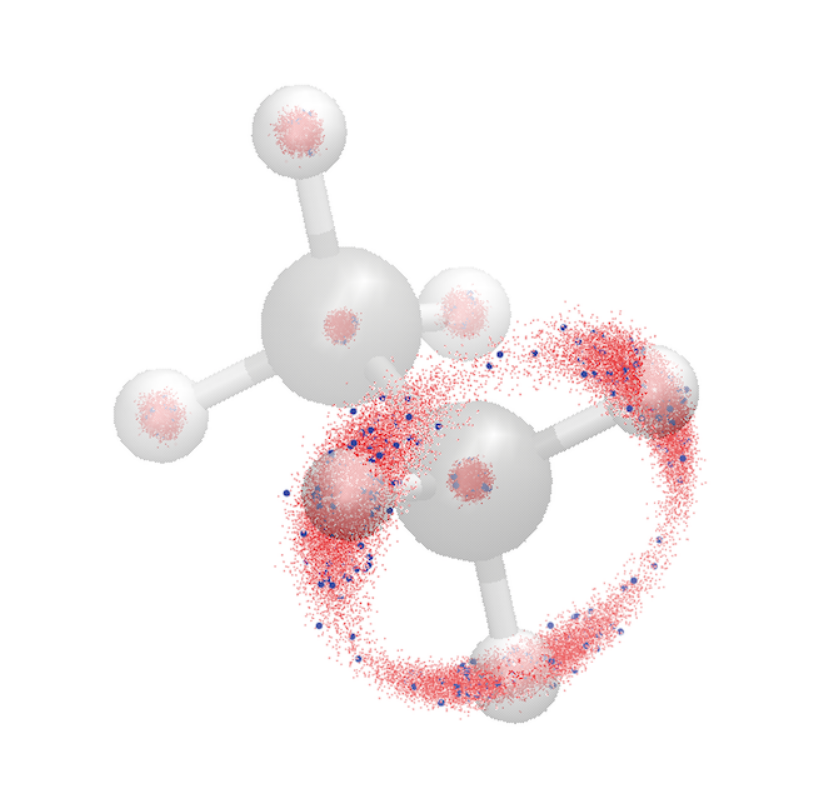}
\caption{\label{fig:ethane-dataset} The extent of the ethane conformers generated by MD (red points). K-means sampling is used to select 2,000 representative points. Test points from an independent trajectory are in blue.}
\end{figure}

For benzene and ethane, conformers were generated from isothermal molecular dynamics (MD) trajectories. The range of atomic positions from combined 1~ns 300~K and 350~K trajectories is shown in Fig.~\ref{fig:benzene-dataset} for benzene and Fig.~\ref{fig:ethane-dataset} for ethane after snapshots are aligned to a reference molecule.

For malonaldehyde, the classical MD trajectories include 0.5~ns for each tautomer at 300~K and 350~K. Resulting conformers used to create the K-means sampled training set are shown as red points in Fig.~\ref{main-fig:malon-dataset} of the main text. The test set to evaluate the energy error is taken from an ab initio MD trajectory at 300~K.  The ML-HK model is also used to \textit{generate} an MD trajectory using a finite difference method to calculate atomic forces at each timestep.  A displacement of $\epsilon$~=~0.001~{\AA}  was chosen to maintain energy conservation during the MD simulation using the Atomistic Simulation Environment (ASE) \cite{BJ02}.  A Langevin thermostat with a friction coefficient of 0.01 atomic units (0.413 fs$^{-1}$) was selected to reproduce the fluctuations in atomic coordinates observed for the trajectory generated using DFT (see Table \ref{tab:malon-rmsd}).  The ML-HK model generates a trajectory that visits molecular conformations at the extremes of the classically-sampled training set, with predicted energies lower than those calculated using DFT directly (see Fig.~\ref{fig:malon-energies}). The largest predicted energy errors are observed for these high-energy conformers. However, the calculated forces are sufficiently large to bring the atoms back toward their equilibrium positions, resulting in a stable molecular trajectory.

\begin{figure}
\includegraphics{malonaldehyde_300Khybrid-350Khybrid_mltest_E_vs_deltaE.pdf}
\caption{\label{fig:malon-energies} Total energy errors from ML-HK generated trajectory snapshots for each relative energy value. The largest energy errors are for high-energy conformations at the extremes of the classical training set coordinates.}
\end{figure}

\begin{table}[h]
  \begin{ruledtabular}
    \begin{tabular}{c|cccc}
    \multirow{1}{*}[-0.6em]{\parbox{1.8cm}{MD Trajectory}} & \multicolumn{4}{c}{Atom Type} \\
    \noalign{\smallskip}
     & C & O & H (-CH) & H (-OH) \\
    \noalign{\smallskip}
    \colrule
    \noalign{\smallskip}
     DFT & 0.052 & 0.076 & 0.166 & 0.289 \\
    \noalign{\smallskip}
     ML-HK & 0.051 & 0.094 & 0.171 & 0.242 \\
     \end{tabular}
\end{ruledtabular}
\caption{\label{tab:malon-rmsd}
RMSD (\AA) for malonaldehyde during 2~ps MD simulations relative to the average coordinates of the two optimized enol tautomers.
}
\end{table}

\section{Sampling}
For $\mathsf{H_2}$, since there is only one atomic distance to adjust, we take the $M$ equi-distant points in the parameter range and for each of these points select the training point that is closest.

For larger molecules with more parameters ($\mathsf{H_2O}$, Benzene, Ethane, Malonaldehyde) we also want to cover the conformer space in a way that all conformers are relatively close to at least one training point.

Assuming $p_i$ are the parameters of conformer $i$ and $i\in \tilde{P}_j$ if and only if $\tilde{p}_j$ is closest to $p_i$, we want to find $\tilde{p}_j$, $j=1\dots M$ that minimize
\begin{align} \label{kmeans_criterion}
  \sum_{j=1}^M \sum_{i\in \tilde{P}_j} ||\tilde{p}_j - p_i||^2.
\end{align}
K-means\citep{kmeans} solves this problem for continuous $\tilde{p}_j$. However, since K-means returns only locally optimal solutions, we rerun the algorithm 50 times and select the solution which minimizes Eq.~\ref{kmeans_criterion}. We choose the points $p_i$ closest to each $\tilde{p}_j$ as training points.

\section{DFT convergence}
For our 3-D DFT calculations in Quantum Espresso\citep{QE2009}, we center a water molecule in a cubic cell and converge three variables: the kinetic energy cutoff for wavefunctions \verb|ecutwfc| in steps of 10 Ry, the kinetic energy cutoff for charge density and potential \verb|ecutrho| in steps of 40 Ry, and the cell dimension \verb|celldm| in steps of 1 bohr. We increase parameters until increasing any parameter does not change the equilibrium position total energy by more than 0.01 kcal/mol for $\mathsf{H_2O}$. We end up with \verb|ecutwfc| of 90 Ry, \verb|ecutrho| of 360 Ry, and \verb|celldm| of 20 bohr, which are used for all other molecules in this work.

\section{Logic of Density Functional Theory (DFT)}
Within the Born-Oppenheimer approximation in non-relativistic quantum mechanics, and using atomic units, the Hohenberg-Kohn paper\citep{HK64} laid the theoretical framework of all modern DFT\@.  The first statement is that the mapping

\begin{equation}
  v(\br) \longleftrightarrow \n(\br)
\end{equation}
is one-to-one, i.e., at most one potential can give rise to
a given ground-state density, even in a quantum many-body problem,
for given interaction among particles and statistics (i.e., fermions
or bosons).  A follow-up claim is that
the ground-state energy of an electronic system can be found
from

\begin{equation}
  E[v] = \min_{\n} \left\{F[\n] + \int d^3 r \, n(\br)v(\br)\right\}
\end{equation}
where $F[\n]$ is a density functional containing all many-body
effects.  The minimizing density is the solution to the Euler
equation:

\begin{equation} \label{eq:euler}
  \frac{\delta F}{\delta \n(\br)} + v(\br) = \mathrm{const}
\end{equation}
It is the direct map between densities and potentials that
we machine-learn in this paper.  We call it the HK density map, $\n[v](\br)$.

The KS scheme avoids direct approximation of $F$ by imagining
a fictitious system of non-interacting electrons with the
same density as the real one\citep{KS65}.
The KS equations are:

\begin{equation} \label{eq:KS}
\left\{ -\frac{1}{2} \nabla^2 + v\s(\br)\right\} \phi_i(\br) = \epsilon_i \phi_i(\br)
\end{equation}
where $\epsilon_i$ are the KS eigenvalues and $\phi_i$ the KS orbitals.

\begin{equation} \label{eq:vs}
v\s(\br) = v(\br) + v\hartree(\br) + v\xc(\br)
\end{equation}
where $v\hartree(\br)$ is the Hartree potential and $v\xc(\br)$ is the exchange-correlation potential.
The true energy of the system is then reconstructed from the
self-consistent density $\n(\br)=\sum_i |\phi_i(\br)|^2$ via

\begin{equation} \label{eq:E}
E[\n] = T\s[\n] + U[\n] + \int d^3 r \, n(\br)v(\br) + E\xc[\n]
\end{equation}
where $T\s[\n]$ is the kinetic energy of the non-interacting electrons and
$U[\n]$ is the Hartree energy. $E\xc[\n]$ is the exchange-correlation (XC)
energy and implicitly defined by Eq.~\ref{eq:E}.
Most calculations\citep{PGB15} use simple
approximations that depend only on the density and its
gradient to determine $E\xc$, called generalized gradient
approximations, or replace a fixed fraction of the approximate
exchange with the exact exchange from a Hartree-Fock
calculation (called a hybrid).
Requiring the XC potential to be the functional derivative of
$E\xc$ ensures that the self-consistent solution of Eq.~\ref{eq:KS}
minimizes the energy of Eq.~\ref{eq:E} for the given $v(\br)$ and
$E\xc[\n]$.
\bibliography{bypassing-ks-with-ml.bib}